%% file: sample-authordraft.tex
\documentclass[sigconf]{acmart}
\AtBeginDocument{%
  \providecommand\BibTeX{{%
    \normalfont B\kern-0.5em{\scshape i\kern-0.25em b}\kern-0.8em\TeX}}}

\copyrightyear{2025}
\acmYear{2025}
\setcopyright{cc}
\setcctype{by}
\acmConference[CHI '25]{CHI Conference on Human Factors in Computing Systems}{April 26-May 1, 2025}{Yokohama, Japan}
\acmBooktitle{CHI Conference on Human Factors in Computing Systems (CHI '25), April 26-May 1, 2025, Yokohama, Japan}
\acmDOI{10.1145/3706598.3713546}
\acmISBN{979-8-4007-1394-1/25/04}

\acmSubmissionID{4471}

\usepackage{siunitx}
\usepackage{caption}
\usepackage{subcaption}
\usepackage[normalem]{ulem}
\usepackage{float}

\begin{document}

%%
%% The "title" command has an optional parameter,
%% allowing the author to define a "short title" to be used in page headers.
\title[Design Patterns for the Common Good]{Design Patterns for the Common Good: Building Better Technologies Using the Wisdom of Virtue Ethics}

%%
%% The "author" command and its associated commands are used to define
%% the authors and their affiliations.
%% Of note is the shared affiliation of the first two authors, and the
%% "authornote" and "authornotemark" commands
%% used to denote shared contribution to the research.
\author{Louisa Conwill}
\email{lconwill@nd.edu}
\orcid{0009-0001-7116-266X}
\affiliation{%
  \institution{University of Notre Dame}
  \city{Notre Dame}
  \state{Indiana}
  \country{USA}
}

\author{Megan K. Levis}
\email{mlevis@nd.edu}
%\orcid{}
\affiliation{%
  \institution{University of Notre Dame}
  \city{Notre Dame}
  \state{Indiana}
  \country{USA}
}

\author{Karla Badillo-Urquiola}
\email{kbadill3@nd.edu}
%\orcid{}
\affiliation{%
  \institution{University of Notre Dame}
  \city{Notre Dame}
  \state{Indiana}
  \country{USA}
}

\author{Walter J. Scheirer}
\email{wscheire@nd.edu}
%\orcid{}
\affiliation{%
  \institution{University of Notre Dame}
  \city{Notre Dame}
  \state{Indiana}
  \country{USA}
}

%%
%% By default, the full list of authors will be used in the page
%% headers. Often, this list is too long, and will overlap
%% other information printed in the page headers. This command allows
%% the author to define a more concise list
%% of authors' names for this purpose.
\renewcommand{\shortauthors}{Conwill et al.}

%%
%% The abstract is a short summary of the work to be presented in the
%% article.
\begin{abstract}
Virtue ethics is a philosophical tradition that emphasizes the cultivation of virtues in achieving the common good. It has been suggested to be an effective framework for envisioning more ethical technology, yet previous work on virtue ethics and technology design has remained at theoretical recommendations. Therefore, we propose an approach for identifying user experience design patterns that embody particular virtues to more concretely articulate virtuous technology designs. As a proof of concept for our approach, we documented seven design patterns for social media that uphold the virtues of Catholic Social Teaching. We interviewed 24 technology researchers and industry practitioners to evaluate these patterns. We found that overall the patterns enact the virtues they were identified to embody; our participants valued that the patterns fostered intentional conversations and personal connections. We pave a path for technology professionals to incorporate diverse virtue traditions into the development of technologies that support human flourishing.
\end{abstract}

%%
%% The code below is generated by the tool at http://dl.acm.org/ccs.cfm.
%%


%%
%% Keywords. The author(s) should pick words that accurately describe
%% the work being presented. Separate the keywords with commas.

%\received{19 January 2023}
%\received[revised]{12 March 2009}
%\received[accepted]{5 June 2009}

%%
%% This command processes the author and affiliation and title
%% information and builds the first part of the formatted document.
\maketitle

\section{Introduction}
How to live with technology well, and how to build technologies that foster our well-being rather than cause harm, are key questions of Human-computer Interaction (HCI). \textit{Virtue ethics} is a philosophical tradition that can help answer these questions. Virtue ethics is one of the three main normative frameworks for ethics in philosophy. In contrast to consequentialism and deontology, the other two primary frameworks in moral philosophy, virtue ethics focuses on the cultivation of virtues --- traits of excellent character --- rather than following a set of moral rules in its determination of what is good. Virtue ethics asks, ``what are the goods in life that we should aim for?'' and, ``what are the qualities of a person that are necessary to act well?'' This relates well with the questions of technology design: what constitutes a good life and what qualities does a technology need to have (or not have) to aim towards those goods? Does the technology foster virtue or vice in its users?

HCI researchers have often handled technologies that have both positive and negative effects by proposing fixes such as adding guardrails to existing designs to minimize their harmful effects~\cite{escher2024hexing, hughes2024viblio, davis2023supporting,zhang2022monitoring,lukoff2021design} or proposing screen time monitors or lockout tools as ways to limit engagement~\cite{hiniker2016mytime,kim2019goalkeeper}. More recently, the Positive Social Technology (\textit{Positech}) movement has emerged as a paradigm shift, focused on building wholly positive technologies from the ground-up rather than fixing or modifying negative ones~\cite{kim2024envisioning}. Similarly, virtue ethics is concerned with striving for excellence rather than simply avoiding problem behaviors, making it a suitable philosophical tradition to pair with the Positech paradigm. Virtue ethics can help inform the goods we should aim for in our positive technology designs.

While scholars have previously considered how virtue ethics can inform technology design, e.g.,~\cite{vallor_2018, gorichanaz2024virtuous, conwill2024virtue}, many of these investigations have remained at a conceptual level (e.g., identifying virtues that would improve social media~\cite{vallor_2018, conwill2024virtue} or search platforms ~\cite{gorichanaz2024virtuous}). We propose an approach for translating this abstract ethical thinking into more structured and re-usable software design recommendations in the form of \textit{design patterns}~\cite{gamma1995pattern}. In our approach we ask: Of the technology designs we encounter every day, which ones are virtuous? Our approach creates a catalog of virtuous design patterns by analyzing existing designs through the lens of a particular virtue tradition and capturing the design patterns that uphold such virtues. The resulting collection of design patterns tells us which designs to continue employing in order to orient our technologies towards the good. Our approach addresses a gap in HCI literature where there is often a disconnect between theoretical ethics and practical implementation. 

To guide our work, we posed the following research questions:
\begin{itemize}
   \item \textbf{RQ1:} \textit{How can we integrate virtue ethics into existing design approaches in order to translate abstract ethical ideas into practical design recommendations?}
    \item \textbf{RQ2:} \textit{Does the proposed process work as intended, identifying design patterns that indeed embody the desired virtues?}
    \item \textbf{RQ3:} \textit{Are the design patterns identified by this process seen as desirable by technology users?}
    \item \textbf{RQ4:} \textit{Would the catalog of design patterns created by this process be adopted in technology development?}
\end{itemize}

To answer \textbf{RQ1}, we adapted existing design methods, including Value Sensitive Design~\cite{friedman2019value} and design patterns, in a new way to produce practical interface recommendations inspired by ethical and religious traditions. As a case study of our process, we identified seven design patterns for social media that uphold Catholic Social Teaching~\cite{compendium2004}: a virtue ethics tradition based on Catholic doctrines concerning human dignity and societal good. To answer \textbf{RQ2}, \textbf{RQ3}, and \textbf{RQ4}, we conducted semi-structured interviews with 24 technologists to gain feedback on the patterns. We found that generally the patterns embodied the principles they sought to characterize and that the patterns have the potential to make a positive impact on the social media landscape if widely adopted. Through the interviews we also identified areas of improvement for the patterns and refined the contexts in which the patterns are most appropriate for use. The contributions of this paper are as follows:
\begin{itemize}
    \item A systematic approach inspired  by virtue ethics for identifying common good-oriented design patterns, thus proposing a practical way to embed ethical values into the design of technology.
    \item Seven design patterns inspired by Catholic Social Teaching to build social media more in accordance with human flourishing, articulating one vision of what such ethically designed online spaces could look like.
    \item The first consideration of how Catholic values can influence design, contributing to the growing field of faith and spirituality within HCI and bringing a new ethical perspective into discussions of technology design.
    \item Implications for designing ethical technologies based on learnings from these design patterns, and a discussion of how these particular patterns might be adopted in the current technology landscape.
\end{itemize}

According to the virtue tradition, human flourishing is constitutive of the good actions we perform~\cite{aristotle_nicomachean_1911}. Because we found that our cataloged design patterns successfully embody their intended virtues, then according to virtue ethics the technologies employing these patterns will be oriented towards the common good: the state of society where every individual can thrive~\cite{Velasquez_Andre_Shanks_Meyer_2018}.

\section{Background}

%Also cite Commit somewhere because it's similar to No Lurking

We situate our research at the intersection of ethical technology design (focusing on ethical social media design in this paper) and philosophy and spirituality in design, highlighting the unique contributions of our work. We also provide overviews of Value Sensitive Design and design patterns, which inform our methods.

\subsection{Ethical Technology Design}
Previous research has considered the ethical design of social media platforms and suggested design implications for greater digital well-being. Much of that work focuses on interventions like browser extensions that turn off harmful features or behavior-changing tools to help users change their relationship to social media~\cite{al2023designing}. However, some researchers have recommended that instead of employing external interventions, the platforms themselves should be redesigned in ways that promote well-being. Previous work has shown this to be a promising path forward for ethical technology design. For instance, a study by Zhang et al., showed that redesigning Twitter (now X) to support user agency is more effective in reducing screen time compared to use of screen time monitoring tools~\cite{zhang2022monitoring}. Similarly, designing for user agency has also been considered in the context of redesigning YouTube~\cite{lukoff2021design}. Finally, reducing meaningless experiences~\cite{lukoff2018makes} and reducing feelings of regret after use~\cite{cho2021reflect} have also guided research of social media design (or redesign) for well-being. Our work aligns with this new path towards designing to promote well-being, and we use virtue ethics to inform the guiding values in the designs.

More recently, HCI scholars have called for a paradigm shift (i.e., \textit{Positech}) for how technologies are built to promote the well-being of users and society rather than simply fixing or modifying the flaws of current platforms to minimize their negative effects~\cite{kim2024envisioning}. This shift seeks to ``pioneer new models'' and build better technologies from the ground-up ~\cite{kim2024envisioning}. To address the call for more wholly positive technologies, we propose an approach that integrates philosophical and spiritual traditions into the ethical design process.

\subsection{Philosophy and Spirituality in Technology Design}
There has been growing interest in HCI to consider how philosophical, religious, or spiritual traditions might relate to technology design~\cite{rifat2022integrating}. For example, Escher and Banovic's  work draws from Greek mythology in calling for new forms of resistance against oppressive technologies~\cite{escher2024hexing}. They propose design modifications that resist harmful patterns baked into already-existing technologies.

Other work has imagined technology design inspired by religious traditions. A recent ACM \textit{Interactions} issue dedicated to spirituality in design included articles on how Judaism~\cite{hammer2022individual}, Christianity~\cite{hiniker2022reclaiming}, and religiosity at large~\cite{toyama2022technology, naqshbandi2022making} could inform technology design. For example, Hiniker and Wobbrock~\cite{hiniker2022reclaiming} proposed that technology designed with Christian principles in mind would promote relationship with God, others, and the environment, and would be at odds with the attention economy --- the primary economic model for many of our technologies today. Similarly, Hammer and Reig discussed how Jewish conceptions of \textit{obligations} can provide a novel conceptual approach and overarching design implications for online speech~\cite{hammer2022individual}. Additionally, Toyama discussed how designing technologies to foster the inner attention valued by religious traditions could ``counter technological excess and corruption''~\cite{toyama2022technology}. Finally, Naqshbandi et al. argued that capturing faith, religion, and spirituality-based motives is important for advancing design justice and getting a more complete picture of human motives and conduct~\cite{naqshbandi2022making}. Beyond the \textit{Interactions} issue, other HCI work has demonstrated that self-tracking apps built with Buddhist principles enable users to develop better emotional and ethical awareness as well as resist technology addiction compared to standard self-tracking apps~\cite{mcguire2020buddhist}.

%Value-Based Engineering is a design approach that draws from Value Sensitive Design, but incorporates philosophical sources. Specifically, Value-Based Engineering relies on guiding questions from the ethical traditions of utilitarianism, virtue ethics, and duty ethics~\cite{spiekermann2020value}. Though Value-Based Engineering provides a process for translating ethical values into system requirements, our work differs from Value-Based Engineering in that Value Based Engineering focuses more on incorporating values into a specific product rather than creating flexible patterns to be used as building blocks for future designs.

All of the works mentioned argue that philosophical or religious traditions should inform technology design. However, most of the proposed design approaches remain at theoretical suggestions. In contrast, our approach allows for principles from diverse ethical traditions to be translated into practical design recommendations.

\subsection{Value Sensitive Design}
One established framework for building more human-centered technologies is Value Sensitive Design (VSD). VSD has grown to encompass a variety of methods that seek to incorporate human values into technology design~\cite{friedman2013value}. The varied methods of VSD rely on empirical studies with stakeholders to identify the values to prioritize.

VSD has been critiqued for the way it identifies the values themselves. Manders-Huits argues that the VSD method of asking stakeholders to provide values does not differentiate between descriptive (provided by stakeholder desires) and normative (what stakeholders \textit{should} want based on some particular ethical framework) values, and is thus unable to distinguish ``genuine moral values from mere preferences, wishes and whims of those involved in the design process.''~\cite{manders2011values} Manders-Huits, along with Jacobs and Huldtgren, argue that Value Sensitive Design should be used with an explicit ethical framework~\cite{manders2011values, jacobs2021value}. If we rely only on what individual stakeholders can come up with on their own, we may miss opportunities to incorporate further wisdom and values into our designs. 

Even so, VSD provides an established and rigorous framework for incorporating values into technology design. As such, our work draws inspiration from VSD methods, but addresses its limitations by incorporating moral wisdom from virtue ethics traditions into the technology design process. Many of these traditions have enduring significance and resonate with many as their guiding values.

The methods of VSD are intended to be used iteratively, integrating conceptual, empirical, and technical investigations~\cite{friedman2013value}. We adopted this tripartite method, specifically inspired by Friedman et al.'s example of VSD for cookies and informed consent in web browsers. This example begins with a conceptual investigation of relevant values, then moves to the development of technical mechanisms to support those values, then to empirical validation of the technical work in light of the conceptual investigations, and finally to the refinement of the technical mechanisms~\cite{friedman2013value}.

One previous work by Reijers and Gordijn incorporates virtue ethics into VSD~\cite{reijers2019}. However, it focuses on the example of military drones and considers the socio-technical systems surrounding military drone usage rather than the designs of the drones themselves. Our adaptation of VSD to incorporate virtue ethics is better suited for the context of UI/UX design and other software matters. Additionally, the method proposed by Reijers and Gordijn relies on Aristotelian virtues, while our proposed method is flexible to incorporate virtues from any tradition.

\subsection{Design Patterns}
% https://wiki.hackerspaces.org/Design_Patterns
In this paper, we propose a process for identifying virtuous design patterns through the lens of virtue ethics traditions. Design patterns are formally documented, re-usable solutions to recurring problems in design~\cite{borchers2000pattern}. If cataloged and disseminated well, design patterns can prevent technologists from reinventing the wheel when faced with tough design problems~\cite{detweiler2012value}. The idea of a design pattern was first introduced in the sixties in the realm of architecture by Christopher Alexander~\cite{alexander1977pattern}. In the nineties, the concept of design patterns began to be used in software engineering~\cite{gamma1995pattern} and then in HCI~\cite{borchers2000pattern}. Design patterns come about organically through re-use of designs that proved effective in the past. Design patterns can be formally documented when recurring patterns of designs are employed across technologies~\cite{gamma1995pattern}. In software engineering, design patterns capture solutions to problems in object-oriented programming, and in user experience, design patterns capture solutions to usability problems. In this paper, we seek to document interface design patterns that capture solutions to problems of virtuous technology usage. When we document virtuous design patterns, we not only look for recurring patterns of designs across technologies but also that those designs uphold particular virtues. While design patterns typically arise as intentional solutions to the problems they solve, the design solutions we document in this paper were not necessarily intended by their designers to promote virtue. More likely, the designs were intended to solve problems of usability, and they incidentally promote certain virtues. 

Similar to our work, Detweiler et al. combined the methods of Value Sensitive Design with design pattern generation~\cite{detweiler2012value}. The authors use a conceptual analysis of pervasive healthcare technologies to identify recurrent issues and stakeholder values, and then create design patterns for more values-based designs. The first and most notable difference with our work is that our work incorporates time-tested virtues into the process, whereas Detweiler et al. take a more classic VSD approach, identifying values from stakeholders. Second, Detweiler et al. create design patterns from only a conceptual investigation of the values in their given technological context, whereas our work catalogs design patterns from a technical investigation of existing designs. In this way, our work uncovers existing design patterns and retains the ones that are virtuous, whereas Detweiler et al. aims to create new designs (which they acknowledge is not in the spirit of the ``time-tested'' nature of design patterns, and that their ``patterns'' should be considered more as proofs-of-concept).

Design patterns for social media exist~\cite{Crumlish_Malone_2015}~\cite{Hussein_Alaa_Hamad_2011}. However, while these design patterns capture the key building blocks for how to build easy-to-use social media applications, these patterns do not make any claims about the ethics or digital well-being they promote. (One study of designs to decrease screen time on Twitter recommends these suggestions to promote digital well-being be cataloged as design patterns, but to our knowledge they have not been~\cite{zhang2022monitoring}.) In contrast, this work documents design patterns for social media that have a particular ethical backing.

\section{Theory}
The theoretical underpinning of our work is \textit{virtue ethics}, and the virtue tradition we use in our proof of concept is \textit{Catholic Social Teaching}. We explain both in more detail below.

\subsection{Virtue Ethics}
Virtue ethics is a philosophical tradition that aims for the common good through the promotion of virtues or good habits. As opposed to other ethical approaches that focus on how to choose rightly when confronted with moral decisions, virtue ethics instead asks, ``what qualities of a person are necessary to act well?" and ``what constitutes a good life?'' When thinking about how virtue ethics applies to technology design, the mindset is shifted from setting up rules to limit or avoid problem behaviors in technologies and towards considering what the goods in life are and how the goods of technology can fit in with the goods in life. In the virtue tradition, good activity is constitutive of human flourishing; in other words, every person performing good activities or practicing virtues helps to achieve the common good. So, if a particular technology fosters virtue, then according to the virtue tradition it is oriented towards the common good.

Virtue ethics is not one tradition but rather an approach that encompasses many traditions, including Aristotelian philosophy, Buddhism, Confucianism, Hinduism, and the Abrahamic religions. Each of these traditions has different sets of virtues that are prioritized as necessary for achieving human flourishing, often based on the cultural context in which that tradition was developed. However, some virtues are universal across all, or most, virtue traditions~\cite{macintyre1981after}.

There are different ways to incorporate virtue ethics into the technology design process. One way is to focus on fostering virtues among individuals and groups involved in the development, implementation, and use of technological systems. The other is to consider virtues in terms of the systems themselves. How can a system foster particular virtues in its users based on its design? What qualities (or ``virtues'') does a system need to possess to make it an excellent system? In this work we address the latter two questions.

The philosopher Shannon Vallor popularized considering the ethics of emerging technologies in terms of virtue ethics through her 2016 book \textit{Technology and the Virtues}~\cite{vallor_2018}. In this book, Vallor analyzes the virtues from the Aristotelian, Buddhist, and Confucianist virtue traditions to identify which are important for living well with technology today. Vallor then considers the ethics of technologies including social media, surveillance, robots, and human enhancement in light of these virtues. Inspired by Vallor, a small number of other scholars have considered how virtue ethics can inform technology design; for example, using virtue ethics to consider the intellectual virtues required for virtuous online search, and proposing high-level design implications from these virtues~\cite{gorichanaz2024virtuous}. To our knowledge, this paper is the first to recommend designs that align with the virtue ethics tradition at the level of specificity of design patterns.

\subsection{Catholic Social Teaching}
As a case study of our proposed approach we will identify design patterns that align with the Catholic tradition. Specifically, the virtues we will consider in our design pattern identification are the main principles of Catholic Social Teaching: the area of Catholic doctrine concerned with upholding human dignity and the common good in society.

Catholic Social Teaching was initially developed as a response to the societal injustices caused by the Industrial Revolution. As the doctrine of Catholic Social Teaching has evolved over time, it has continued to respond to the societal ills caused by new technologies, including the mass media and nuclear weapons of the 1960s, and more recently, the Internet. In other words, the particular virtues and values emerging from Catholic Social Teaching specifically target societal ills caused by technology. As such, it is fitting to continue to use Catholic Social Teaching to think about how to design and use technology well~\cite{conwill2024virtue}. The Catholic Church's guidance on technology is also not unsolicited: many Silicon Valley technology executives have sought out ethical advice from the Vatican~\cite{germain2023}.  Additionally, in 2024 Pope Francis was the first-ever pontiff to address a G7 summit, speaking about AI ethics~\cite{winfield2024}. This further demonstrates the desire for Catholic perspectives in technology ethics. Scholars are now exploring how Catholic Social Teaching can inform the ethics of newer digital technologies. This includes Conwill et al.'s analysis of Catholic Social Teaching and social media~\cite{conwill2024virtue} and Gaudet et al.'s consideration of its applications to artificial intelligence~\cite{gaudet2024}. Unlike other aspects of Catholic doctrine that are meant primarily for those who profess the Catholic faith, Catholic Social Teaching is meant to draw any person concerned with social justice into conversation about how to build a better world.

While we hope that our proposed process will be employed with virtues from various virtue ethics traditions, we selected Catholic Social Teaching for our proof of concept because of the aforementioned particular qualities that make it highly relevant to technology and accessible to people of all backgrounds.

\section{Approach: Virtue-Guided Technology Design}\label{sec:approach}

\begin{figure*}
    \centering
    \includegraphics[width=\linewidth]{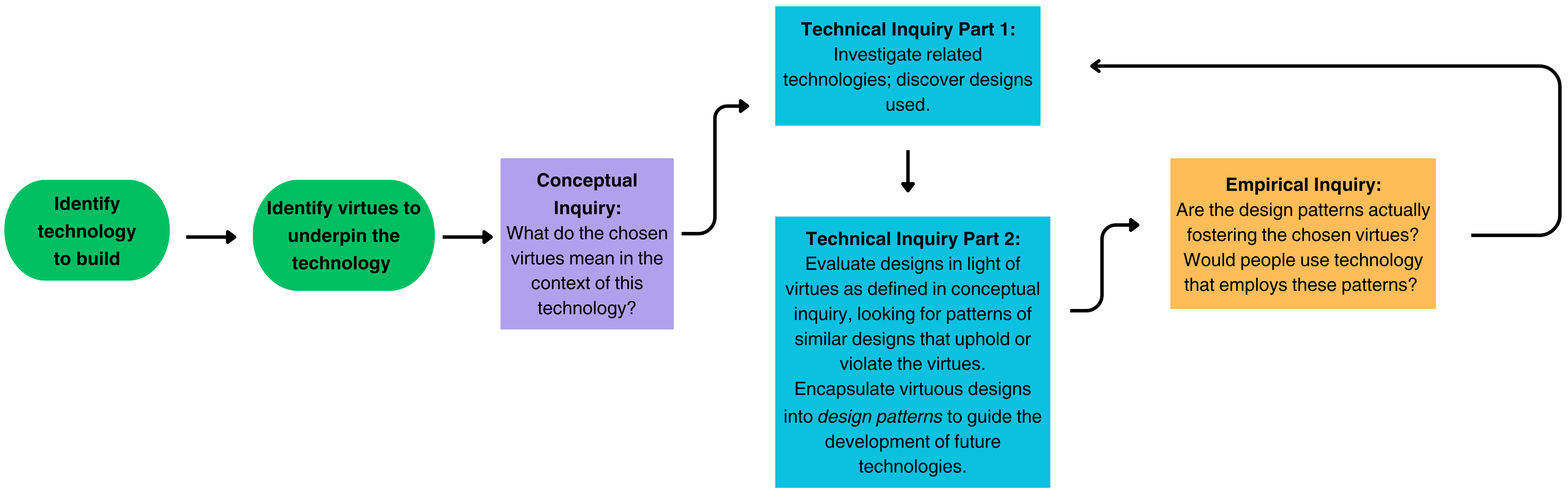}
    \caption{Process diagram describing our proposed process for Virtue-Guided Technology Design. The process begins with identifying a technology to build and the virtues that will underpin it, then moves to a conceptual inquiry of what the virtues mean in the context of the chosen technology, a two-part technical inquiry to discover and catalog the design patterns that uphold such virtues, and an empirical inquiry to validate the process.}
    \Description{On the left is a green bubble that says, "identify technology to build." There is an arrow from that bubble to another green bubble that says, "identify virtues to underpin the technology." There is an arrow from that bubble to a purple box that says, "Conceptual Inquiry: What do the chosen virtues mean in the context of this technology?" From that box there is an arrow to another box that says, "Technical Inquiry Part 1: Investigate related technologies; discover designs used." From that box there is an arrow to another blue box that says, "Technical Inquiry Part 2: Evaluate designs in light of virtues as defined in conceptual inquiry, looking for patterns of similar designs that uphold or violate the virtues. Encapsulate virtuous designs into design patterns to guide the development of future technologies." From that box there is an arrow to a yellow box that says, "Empirical Inquiry: Are the design patterns actually fostering the chosen virtues? Would people use technology that employs these patterns?" The empirical inquiry box has an arrow pointing back to the technical inquiry part 1 box.}
    \label{fig:process_diagram}
\end{figure*}

We propose an approach that we call \textit{Virtue-Guided Technology Design} for using virtues to inspire a catalog of common good-oriented design patterns. Our approach investigates various technologies to discover common designs, determines how these designs uphold or violate particular virtues, and documents design patterns that encapsulate how such designs are solutions to problems of virtuous technology usage.

Our proposed approach is inspired by two previous works. The first is Friedman et al.'s example of VSD for cookies and informed consent in web browsers~\cite{friedman2013value}. Second, our incorporation of design pattern generation into VSD methodology is inspired by the work of Detweiler et al., who used VSD to create design patterns to support positive human values in pervasive healthcare technologies~\cite{detweiler2012value}. However, our work differs from Detweiler et al. through our incorporation of virtue ethics, through our inclusion of technical and empirical inquiries, and through our placement of design pattern generation at the stage of the technical inquiry rather than at the conceptual inquiry. An overview of our proposed approach can be see in Figure \ref{fig:process_diagram}.

Inspired by VSD~\cite{friedman2019value}, our approach begins with identifying two fundamental motivators: 1) the values to design for and 2) the type of technology to be built. In our approach, selecting the values --- or virtues --- to design for first requires identifying a virtue ethics tradition to draw from. As an example, a designer may choose to design a video streaming platform that embodies the Aristotelian virtue of temperance: the virtue that counters overindulgence~\cite{Aristotle_2000}. Once the virtues to design for are identified, the \textit{Conceptual Inquiry} begins. 

\textbf{Conceptual Inquiry:} This stage entails considering how these virtues may play out in the context of the technology one desires to design. According to Friedman et al.~\cite{friedman2013value}, ``Conceptual investigations do not by themselves involve costly empirical analyses, but instead thoughtful consideration of how stakeholders might be socially impacted by one's technological designs.'' In our example, the designer may conclude after thoughtful consideration that a video streaming platform embodying temperance would not encourage users to watch content for hours on end.

\textbf{Technical Inquiry:} After working definitions for how the chosen values play out in an online context are established, the process moves on to the \textit{Technical Inquiry}. First, identify already-existing technologies that are similar to the technology the design patterns are being cataloged for. Then, analyze the technical features of these example technologies, considering how they hold up in light of the conceptual definitions of the identified virtues. This results in a list of features that either promote or hinder the chosen virtues. In our example, the designer might consider Netflix, and note that its autoplay feature violates temperance in the way she identified in the conceptual inquiry.

\textit{Design Patterns} can then be named and described from the list of features that either promote or hinder the chosen virtues. This is similar to the process of how design patterns typically come to be: someone names and describes a repeated design solution that they have observed. However, in this process, we only retain the patterns that uphold the virtues as defined in the conceptual inquiry. Common features that promote the chosen virtues can be encapsulated into design patterns, and in some cases, features that violate the chosen virtues may additionally justify the documentation of designs that do the opposite. In our running example, the designer who observers Netflix's autoplay feature as a violation of temperance may document a design pattern that promotes only playing videos selected by the user, especially if she observes other video platforms that require individual selection over autoplay.

\begin{figure}[b]
    \centering
    \includegraphics[width=0.5\linewidth]{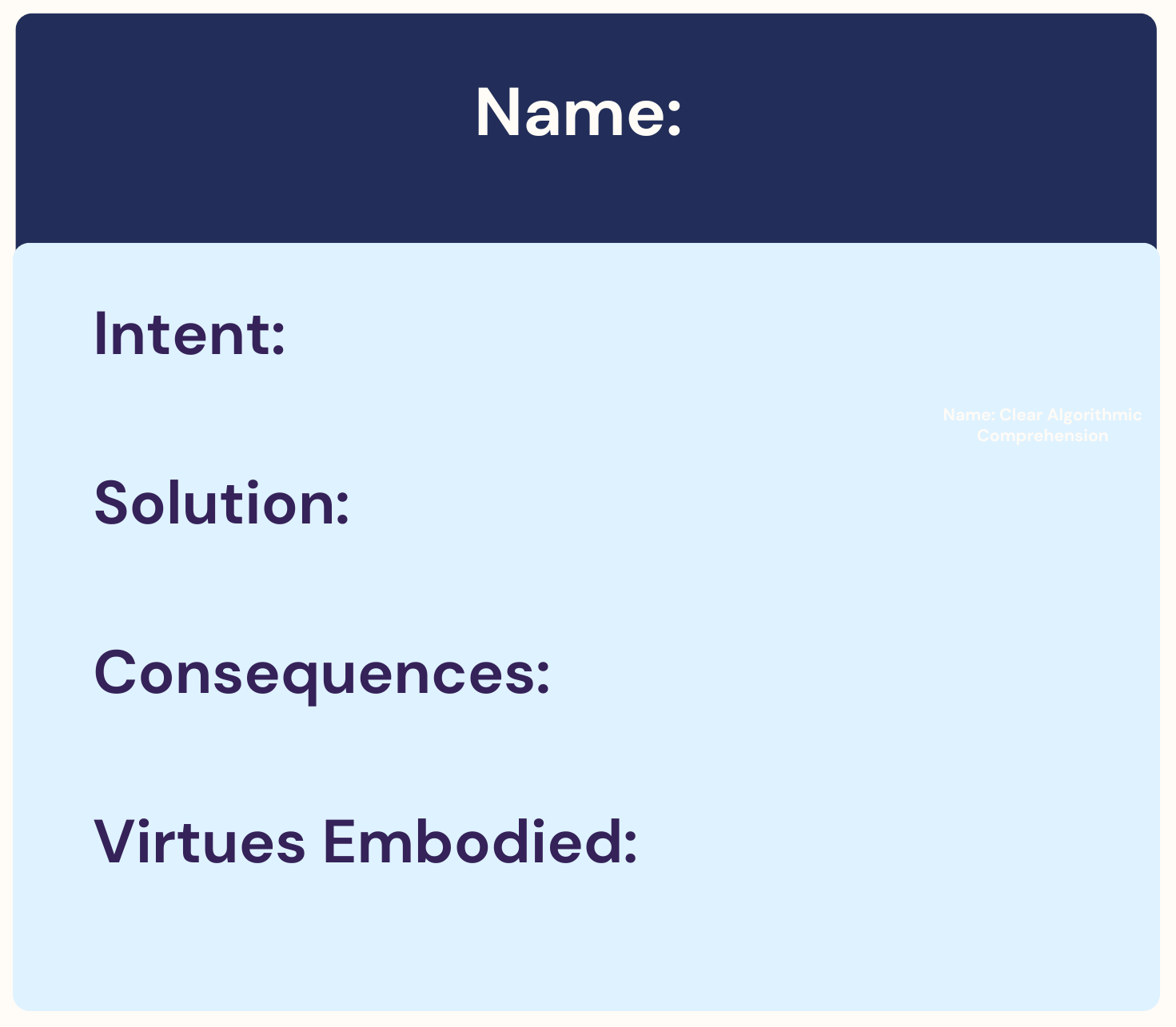}
    \caption{The template for our virtue-inspired design patterns, adapted from \textit{Design Patterns: Elements of Reusable Object-Oriented Software}~\cite{gamma1995pattern}. Each design pattern consists of this information.}
    \Description{A blue box with the following on their own lines: "Name" with a colon, "Solution" with a colon, "Consequences" with a colon, and "Virtues Embodied" with a colon.}
    \label{fig:design_patterns_template}
    \vspace{-10pt}
\end{figure}

The seminal book \textit{Design Patterns: Elements of Reusable Object-Oriented Software} describes a design pattern as having four essential elements: the \textbf{pattern name}, the \textbf{problem}, or the intention behind the pattern and when to apply it, the \textbf{solution}, or the elements that make up the design, and the \textbf{consequences}, or the trade-offs in applying the pattern~\cite{gamma1995pattern}. Detweiler et al.~\cite{detweiler2012value} additionally included the values underpinning the design pattern in their VSD-inspired design patterns. Thus we use the template in Figure \ref{fig:design_patterns_template} for our design patterns.

\textbf{Empirical Inquiry:} Once the design patterns have been documented, the process moves to the \textit{Empirical Inquiry} to validate that the patterns reflect the virtues they were identified to uphold. Empirical inquiries can also investigate the extent to which the patterns would improve user experience if widely, or exclusively, adopted. For this, we recommend semi-structured interviews (inspired by the value-oriented semi-structured interviews from VSD~\cite{friedman2019value}). The empirical studies thus justify the inclusion of each design pattern in the catalog, and can identify areas of opportunity for further design patterns.

\begin{figure*}
    \centering
    \includegraphics[width=0.88\linewidth]{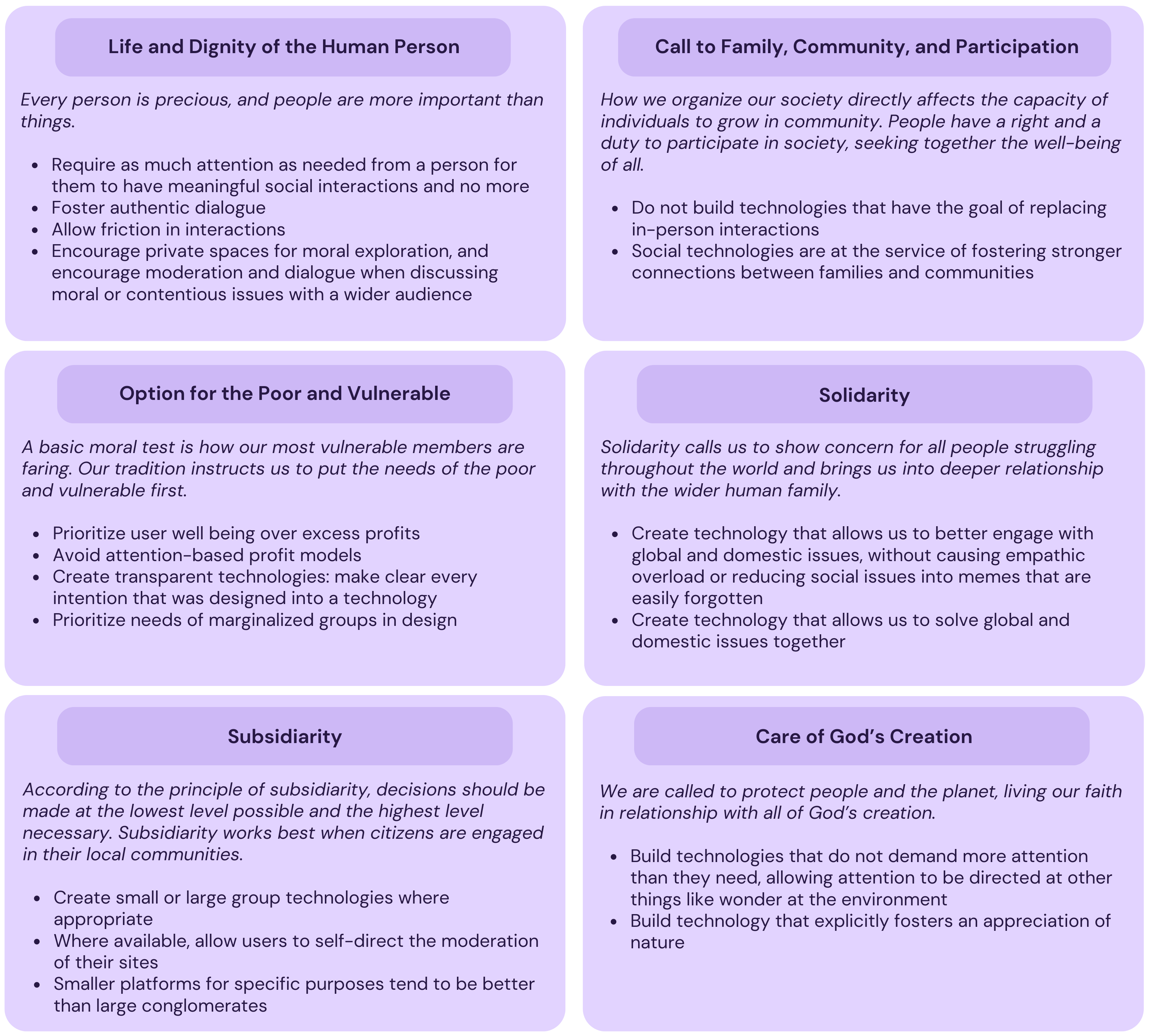}
    \caption{Conceptual Inquiry of what Catholic Social Teaching principles mean in the context of social media platforms, adapted from the book \textit{Virtue in Virtual Spaces}~\cite{conwill2024virtue}.}
    \label{fig:conceptual_inquiry}
    \Description{The six principles of Catholic Social Teaching we consider in this work are listed here with a short definition and bullet points explaining how the principle applies in the context of social media. They are as follows. Principle 1,  Life and Dignity of the Human Person. Definition: Every person is precious, and people are more important than things. Conceptual inquiry: Require as much attention as needed from a person for them to have meaningful social interactions and no more. Foster authentic dialogue. Allow friction in interactions. Encourage private spaces for moral exploration, and encourage moderation and dialogue when discussing moral or contentious issues with a wider audience. Principle 2, Call to family, community, and participation. Definition: How we organize our society directly affects the capacity of individuals to grow in community. Conceptual Inquiry: People have a right and a duty to participate in society, seeking together the well-being of all. Do not build technologies that have the goal of replacing in-person interactions. Social technologies are at the service of fostering stronger connections between families and communities. Principle 3, option for the poor and vulnerable. Definition: A basic moral test is how our most vulnerable members are faring. Our tradition instructs us to put the needs of the poor and vulnerable first. Conceptual Inquiry: Prioritize user well being over excess profits. Avoid attention-based profit models. Create transparent technologies: make clear every intention that was designed into a technology. Prioritize needs of marginalized groups in design. Principle 4, solidarity. Definition: Solidarity calls us to show concern for all people struggling throughout the world and brings us into deeper relationship with the wider human family. Conceptual inquiry: Create technology that allows us to better engage with global and domestic issues, without causing empathic overload or reducing social issues into memes that are easily forgotten. Create technology that allows us to solve global and domestic issues together. Principle 5, subsidiarity. Definition: According to the principle of subsidiarity, decisions should be made at the lowest level possible and the highest level necessary. Subsidiarity works best when citizens are engaged in their local communities. Conceptual inquiry: Create small or large group technologies where appropriate. Where available, allow users to self-direct the moderation of their sites. Smaller platforms for specific purposes tend to be better than large conglomerates. Principle 6, care of God's creation. Definition: We are called to protect people and the planet, living our faith in relationship with all of God’s creation. Conceptual inquiry: Build technologies that do not demand more attention than they need, allowing attention to be directed at other things like wonder at the environment. Build technology that explicitly fosters an appreciation of nature.}
\end{figure*}

\begin{figure*}
    \centering
    \includegraphics[width=\linewidth]{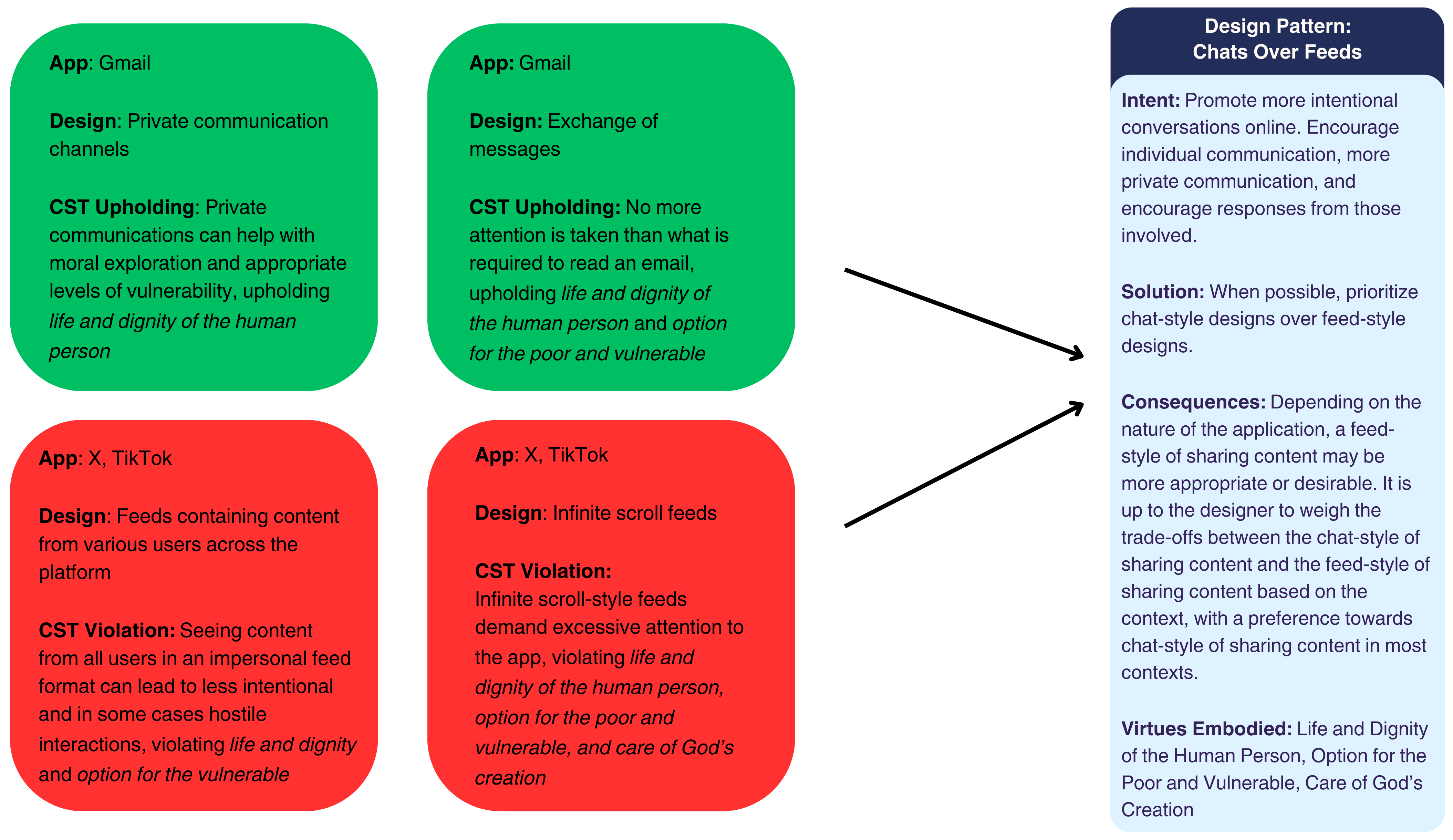}
    \caption{An illustration of the technical inquiry process. When examining the design features that upheld or violated \textit{life and dignity of the human person}, \textit{option for the poor and vulnerable}, and \textit{care of God's creation}, we noticed a pattern: private messages with a finite length in Gmail uphold these principles, and the opposing design of public messages with infinite amounts of content in X and TikTok violate these principles. Considered together, this inspired us to document the ``Chats Over Feeds'' virtuous design pattern.}
    \label{fig:technical_inquiry}
    \Description{On the top left there are two green bubbles. The first includes the following text: App: Gmail. Design: Private communication channels. CST Upholding: Private communications can help with moral exploration and appropriate levels of vulnerability, upholding life and dignity of the human person. The second includes the following text: App: Gmail. Design: Exchange of messages. CST Upholding: No more attention is taken than what is required to read an email, upholding life and dignity of the human person and option for the poor and vulnerable. On the bottom left there are two red bubbles. The first includes the following text: App: X, TikTok. Design: Feeds containing content from various users across the platform. CST Violation: Seeing content from all users in an impersonal feed format can lead to less intentional and in some cases hostile interactions, violating life and dignity and option for the vulnerable. The second includes the following text: App: X, TikTok. Design: Infinite scroll feeds. CST Violation: Infinite scroll-style feeds demand excessive attention to the app, violating care of God’s creation, life and dignity of the human person, and option for the poor and vulnerable. The top and bottom rows of bubbles have arrows to a blue box that says: Design Pattern: Chats Over Feeds. Intent: Promote more intentional conversations online. Encourage individual communication, more private communication, and encourage responses from those involved. Solution: When possible, prioritize chat-style designs over feed-style designs. Consequences: Depending on the nature of the application, a feed style of sharing content may be more appropriate or desirable. It is up to the designer to weigh the trade-offs between the chat-style of sharing content and the feed-style of sharing content based on the context, with a preference towards chat-style of sharing content in most contexts. Virtues Embodied: Life and Dignity of the Human Person, Option for the Poor and Vulnerable, Care of God’s Creation.
}
\end{figure*}

\section{Methods: Catholic Social Teaching-Guided Social Media Design as a Proof Of Concept for Virtue-Guided Technology Design}
As a proof of concept of our proposed approach, we use Virtue-Guided Technology Design to identify design patterns for social media platforms inspired by the virtues of Catholic Social Teaching.

\subsection{Conceptual Inquiry: What Catholic Social Teaching Principles Mean in the Context of Social Media}
The book \textit{Virtue in Virtual Spaces: Catholic Social Teaching and Technology} considers how the virtues, or principles, of Catholic Social Teaching apply in a technological context, especially to social technologies~\cite{conwill2024virtue}. The authors draw on technology ethics literature such as \textit{Technology and the Virtues}~\cite{vallor_2018} and connect these technology ethics concepts to the principles of Catholic Social Teaching. This analysis forms the basis of our conceptual inquiry. \textit{Virtue in Virtual Spaces} identifies eight of the main principles of Catholic Social Teaching: (1) \textit{life and dignity of the human person}; (2) \textit{call to family, community, and participation}; (3) \textit{rights and responsibilities}; (4) \textit{option for the poor and vulnerable}; (5) \textit{dignity of work and rights of workers}; (6) \textit{solidarity}; (7) \textit{subsidiarity}; and (8) \textit{care of God's creation}. Of these eight principles, \textit{dignity of work and rights of workers} and \textit{rights and responsibilities} produced conceptual definitions that pertained only to company organization rather than to technology design. As such, we omitted these two principles from our conceptual investigation. We also omitted any elements of the principles from \textit{Virtue in Virtual Spaces} that pertained to company organization rather than to technology design. The pared-down list of conceptual definitions we considered can be seen in Figure \ref{fig:conceptual_inquiry}.

\subsection{Technical Inquiry: Which Social Media Design Features Uphold or Violate Catholic Social Teaching}

In the technical inquiry, we identified the designs used in various social communications platforms and checked them against the conceptual definitions of Catholic Social Teaching principles to identify patterns of virtuous (or unvirtuous) designs. We analyzed the following social communications platforms in light of the conceptual definitions: X (formerly known as Twitter),  TikTok, BeReal, Gmail, and Discord. We selected these platforms because while they are all social technologies, they encapsulate a wide variety of features and mediums through which communication occurs. First, for each platform we listed out a comprehensive set of features from that platform. Then, for each feature we considered how it holds up in terms of the conceptual definitions of each principle of Catholic Social Teaching as articulated in \textit{Virtue in Virtual Spaces}. Finally, we categorized features by the principles of Catholic Social Teaching they upheld or violated.

\begin{figure*}
    \centering
    \includegraphics[width=\linewidth]{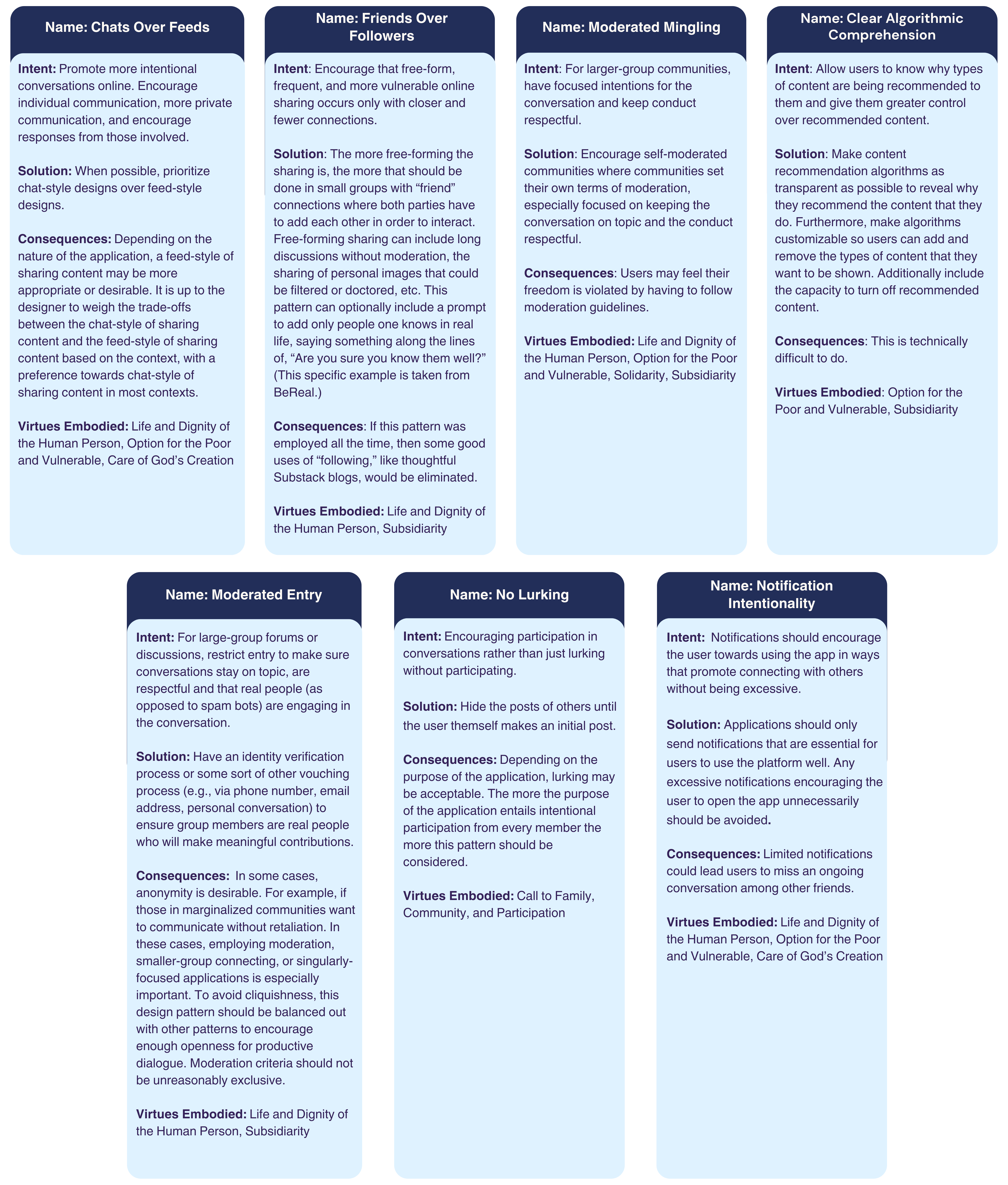}
    \caption{The seven design patterns for social media we documented that uphold the principles of Catholic Social Teaching.}
    \label{fig:design_patterns}
    \Description{The seven design patterns are listed with their name, intent, solution, consequences, and virtues embodied. They are as follows. Pattern 1. Name: Chats Over Feeds. Intent: Promote more intentional conversations online. Encourage individual communication, more private communication, and encourage responses from those involved. Solution: When possible, prioritize chat-style designs over feed-style designs. Consequences: Depending on the nature of the application, a feed style of sharing content may be more appropriate or desirable. It is up to the designer to weigh the trade-offs between the chat-style of sharing content and the feed-style of sharing content based on the context, with a preference towards chat-style of sharing content in most contexts. Virtues Embodied: Life and Dignity of the Human Person, Option for the Poor and Vulnerable, Care of God’s Creation. Pattern 1. Name: Friends Over Followers. Intent: Encourage that free-form, frequent, and more vulnerable online sharing occur only with closer and fewer connections. Solution: The more free-forming the sharing is, the more that should be done in small groups with “friend” connections where both parties have to add each other in order to interact. Free-forming sharing can include long discussions without moderation, the sharing of personal images that could be filtered or doctored, etc. This pattern can optionally include a prompt to add only people one knows in real life, saying something along the lines of, “Are you sure you know them well?” (This specific example is taken from BeReal.) Consequences: If this pattern was employed all the time, then some good uses of “following,” like thoughtful Substack blogs, would be eliminated. Virtues Embodied: Life and Dignity of the Human Person, Subsidiarity/ Pattern 3. Name: Moderated Mingling. Intent: For larger-group communities, have focused intentions for the conversation and keep conduct respectful. Solution: Encourage self-moderated communities where communities set their own terms of moderation, especially focused on keeping the conversation on topic and the conduct respectful. Consequences: Users may feel their freedom is violated by having to follow moderation guidelines. Virtues Embodied: Life and Dignity of the Human Person, Option for the Poor and Vulnerable, Solidarity, Subsidiarity. Pattern 4. Name: Clear Algorithmic Comprehension. Intent: Allow users to know why types of content are being recommended to them and give them greater control over recommended content. Solution: Make content recommendation algorithms as transparent as possible to reveal why they recommend the content that they do. Furthermore, make algorithms customizable so users can add and remove the types of content that they want to be shown. Additionally include the capacity to turn off recommended content. Consequences: This is technically difficult to do. Virtues Embodied: Option for the Poor and Vulnerable, Subsidiarity. Pattern 5. Name: Moderated Entry. Intent: For large-group forums or discussions, restrict entry to make sure conversations stay on topic, are respectful and that real people (as opposed to spam bots) are engaging in the conversation Solution: Have an identity verification process or some sort of other vouching process (e.g. via phone number, email address, personal conversation, etc.) to ensure group members are real people who will make meaningful contributions. Consequences: In some cases, anonymity is desirable. For example, if those in marginalized communities want to communicate without retaliation. In these cases, employing moderation, smaller-group connecting, or singularly-focused applications are especially important. To avoid cliquishness, this design pattern should be balanced out with other patterns to encourage enough openness for productive dialogue. Moderation criteria should not be unreasonably exclusive. Virtues Embodied: Life and Dignity of the Human Person, Subsidiarity. Pattern 6. Name: No Lurking. Intent: Encouraging participation in conversations rather than just lurking without participating. Solution: Hide the posts of others until the user themselves makes an initial post. Consequences: Depending on the purpose of the application, lurking may be acceptable. The more the purpose of the application entails intentional participation from every member the more this pattern should be considered. Virtues Embodied: Call to Family, Community, and Participation. Pattern 7. Name: Notification Intentionality. Intent:  Notifications should encourage the user towards using the app in ways that promote connecting with others without being excessive. Solution: Applications should only send notifications that are essential for users to use the platform well. Any excessive notifications encouraging the user to open the app unnecessarily should be avoided. Consequences: Limited notifications could lead users to miss an ongoing conversation among other friends. Virtues Embodied: Life and Dignity of the Human Person, Option for the Poor and Vulnerable, Care of God’s Creation.}
\end{figure*}

We then documented design patterns based on how we observed patterns of features upholding or violating the principles of Catholic Social Teaching according to their conceptual definitions. An illustration of this process for one design pattern can be seen in Figure \ref{fig:technical_inquiry}; illustrations for the other patterns can be found in the appendix. After generating eleven potential patterns, we had iterative group discussions to choose a subset of patterns that best embodied the principles of Catholic Social Teaching, emphasized interpersonal communication, and/or had the potential to generate thoughtful discussions. A preliminary set of seven patterns were chosen, including: 1) Chats Over Feeds, 2) Friends Over Followers, 3) Moderated Mingling, 4) Clear Algorithmic Comprehension, 5) Moderated Entry, 6) No Lurking, and 7) Notification Intentionality. Additional information about each pattern can be found in Figure \ref{fig:design_patterns}.

\subsection{Empirical Inquiry: Evaluating the Design Patterns}
We conducted an IRB-approved semi-structured interview study to evaluate the design patterns.

\input{participant_table}

\subsubsection{Study Overview} We conducted semi-structured interviews with 24 technology researchers and industry practitioners in person and on Zoom (with cameras turned on), depending on feasibility and the preference of the participant. Participants were first asked to share about their work experience, experience with design and design patterns, and provide basic demographic information. Then, the participants were shown each of the seven design patterns one-by-one. Participants were presented with a shortened version of the design pattern, including the name, intent, and solution. The consequences and virtues embodied were removed from the patterns to not bias participant responses about the positives and negatives of each pattern.

For each pattern, participants were asked to describe how they thought the pattern might have a positive impact on social media, how the pattern might have a negative impact on social media, and whether the pattern would have an overall net positive, net negative, or net neutral impact on the social media landscape if widely adopted. At the end of the interview, participants whose work in industry related to product design, user experience design or research, or front end/mobile engineering were asked additional questions about if the patterns would be valuable in a corporate environment. Of those participants, those who did not have a Christian background were additionally asked if the patterns and the themes of Catholic Social Teaching that inspired them resonated with a non-Christian background.

Before seeing the design patterns, participants were told that every design pattern has inherent pros and cons, that these design patterns are likely not perfect, and in fact some of the patterns were included for the purpose of sparking discussion. This disclaimer was intended to prevent participants from being biased towards providing only favorable feedback on the patterns, due to a desire to support design patterns inspired by their religious beliefs (in the case of Catholic participants), a desire to please the interviewer, or a desire to like the patterns because they are intended to promote social good.

\subsubsection{Participant Recruitment \& Demographics:} We recruited 24 interview participants, at which point little new information was generated from the interviews, reaching data saturation \cite{saunders2018saturation}. Participants were required to be at least 18 years of age and were required to be a developer, researcher, or designer with a background in computer science, software engineering, human-computer interaction, or a related discipline. Participants with these backgrounds were selected because of their familiarity with technology, which would assist in evaluating design patterns. They could additionally rely on their experience to better evaluate the usefulness of particular design patterns in industry settings. Participants were recruited through word of mouth, participant referral, and social media including X and relevant Slack teams. Participants participated in this study on a volunteer basis. At the time of participation, all participants resided in the United States. Table \ref{tab:participant_demographics} summarizes our participant demographics.

\subsubsection{Analysis Approach}
We used reflexive thematic analysis~\cite{braun2019reflecting} to analyze the interview responses. We inductively coded for participants' perspectives on the positive and negative impacts these patterns could have on social media. We then synthesized the codes searching for patterns of values and concerns, combining codes into overarching themes. This resulted in eight overarching themes for the aspects of the patterns participants valued, and eight overarching themes for the concerns participants had about the patterns. The values that our participants saw reflected in the patterns were \textit{meaningful conversation}, \textit{personal connection}, \textit{virtuous conversation}, \textit{not exploitatitve}, \textit{control}, \textit{online safety and privacy}, \textit{transparency}, and \textit{autonomy}. The concerns were \textit{limited learning}, \textit{implementation}, \textit{limited social group}, \textit{challenges of moderation}, \textit{no privacy or positive anonymity}, \textit{increased moral distance}, \textit{forced to share unnecessarily}, and \textit{undesirable self knowledge}.

\section{Results}
In this section, we begin by analyzing whether our design patterns enact the principles of Catholic Social Teaching they were identified to embody. We then consider participant responses to whether the patterns would have a net positive, net negative, or net neutral effect on social media, and if the catalog of patterns would be adopted in industry.

\subsection{Are these patterns enacting the principles of Catholic Social Teaching?}
The participant-identified values and concerns provide insight into the user experience of each pattern. To evaluate how our design patterns would enact the principles of Catholic Social Teaching in their use, we mapped these values and concerns onto embodiments and violations of Catholic Social Teaching principles. We then used these mappings to identify which principles of Catholic Social Teaching each pattern embodies or violates.

\subsubsection{Mapping the Participant-Identified Values to Embodiments of Catholic Social Teaching}\label{sec:embodiment}
\begin{figure*}
    \centering
    \includegraphics[width=\linewidth]{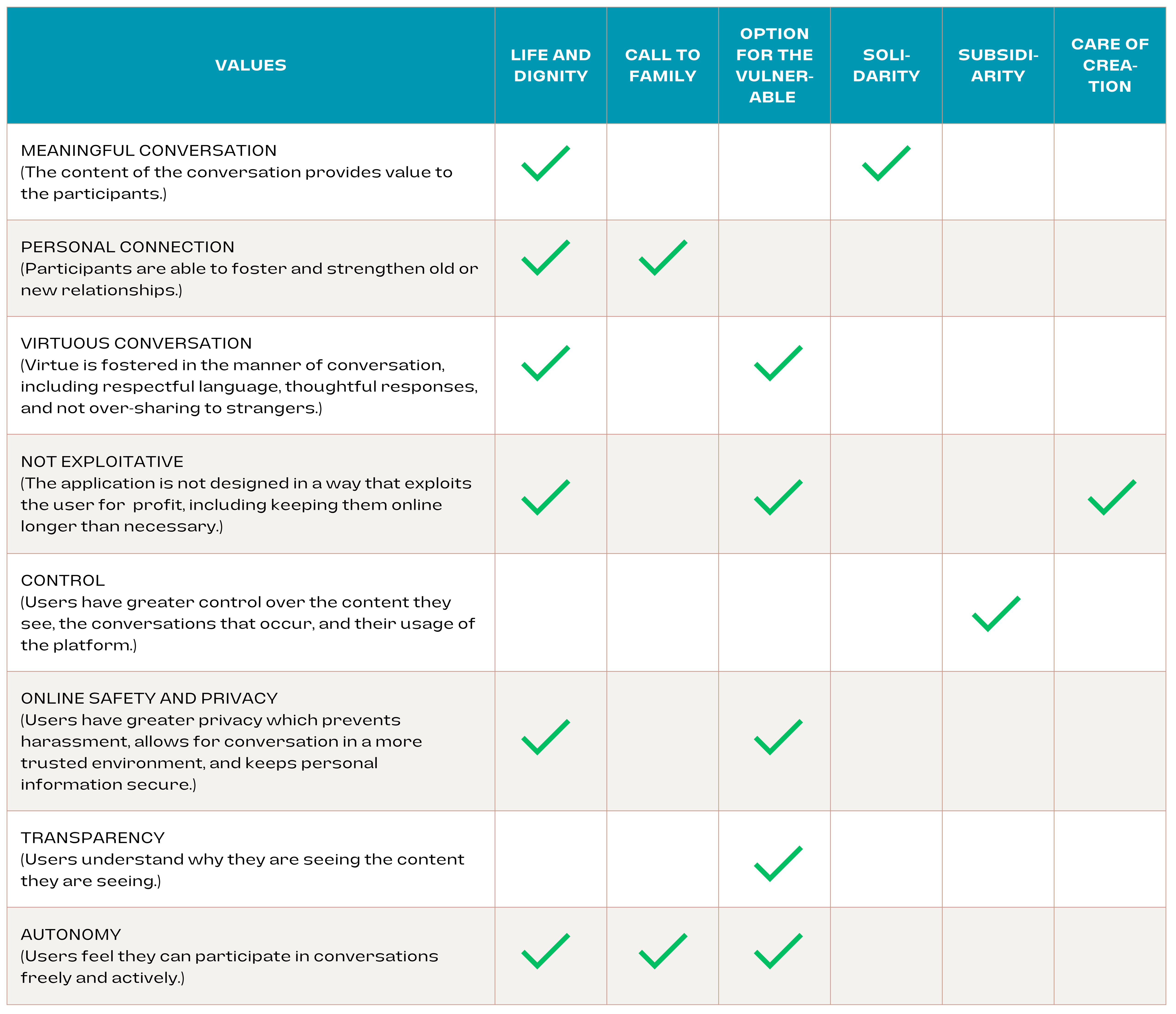}
    \caption{Values that our interview participants identified as being embodied by our design patterns, and which principles of Catholic Social Teaching these values uphold. While all principles were embodied by at least one pattern, life and dignity of the human person and option for the poor and vulnerable were the most embodied by our participant-identified values.}
    \label{fig:upside_themes_to_cst}
    \Description{Comparison chart comparing the participant-identified values our design patterns embody with the principles of Catholic Social Teaching used to inspire the patterns. For each value, the themes of Catholic Social Teaching it embodies are checked off. The values listed down the left side are as follows. Meaningful conversation (The content of the conversation provides value to the participants.) Life and dignity and solidarity are checked. Personal Connection (Participants are able to foster and strengthen old or new relationships.) Life and dignity and call to family are checked. Virtuous Conversation (Virtue is fostered in the manner of conversation, including respectful language, thoughtful responses, and not over-sharing to strangers.) Life and dignity and option for the vulnerable are checked. Not Exploitative (The application is not designed in a way that exploits the user for  profit, including keeping them online longer than necessary.) Life and dignity, option for the vulnerable, and care of creation are checked. Control (Users have greater control over the content they see, the conversations that occur, and their usage of the platform.) Subsidiarity is checked. Online safety and privacy (Users have greater privacy which prevents harassment, allows for conversation in a more trusted environment, and keeps personal information secure.) Life and dignity and option for the vulnerable are checked. Transparency (Users understand why they are seeing the content they are seeing.) Option for the vulnerable is checked. Autonomy (Users feel they can participate in conversations freely and actively.) Life and dignity, call to family, and option for the vulnerable are checked.}
\end{figure*}

We mapped our eight participant-identified values onto the principles of Catholic Social Teaching they correspond to. This mapping can be seen in Figure \ref{fig:upside_themes_to_cst}. The principles of Catholic Social Teaching often overlap and go hand-in-hand: for example, promoting the \textit{option for the poor and vulnerable} implicitly promotes the \textit{dignity of the human person} as well. Because of this, we focused mainly on the explicit promotions of the principles of Catholic Social Teaching, especially as they relate to our conceptual definitions of the principles in light of social media, rather than implicit promotions of the principles.\\

\textbf{Meaningful conversation} (\textit{life and dignity of the human person, solidarity}): Our participants described meaningful conversation as \textit{``spaces for people to engage in the discourse they actually want''} (P23) and \textit{``a communication channel, rather than just broadcasting''} (P16).  The theme embodies comments about the content of the conversation providing value to the participants. The conceptual definition of life and dignity of the human person relates to authentic dialogue and moral exploration, and the conceptual definition of solidarity relates to engaging with the global human family. Both of these can be achieved through meaningful conversation. Meaningful conversation was mentioned for the following patterns: Chats Over Feeds (N=18), Friends Over Followers (N=10), Moderated Mingling (N=17), Moderated Entry (N=12), No Lurking (N=5), and Notification Intentionality (N=1).

\textbf{Personal connection} (\textit{life and dignity of the human person; call to family, community, and participation}): Personal connection relates to participant comments about the ability to foster and strengthen old or new relationships. It was often but not always mentioned alongside meaningful conversation. P21 described this as, \textit{``promoting authentic communication with your genuine friends over this sense of polished facade-building for the world.''} The call to family, community, and participation calls for fostering relationships; life and dignity of the human person calls for authentic dialogue. Both of these are strengthened by personal connections. Personal connection was mentioned for the following patterns: Chats Over Feeds (N=15), Friends Over Followers (N=13), Moderated Mingling (N=3), Moderated Entry (N=11), No Lurking (N=11), Notification Intentionality (N=2).

\textbf{Virtuous conversation} (\textit{life and dignity of the human person, option for the poor and vulnerable}): This value relates to the conduct of participants in conversations, including respectful language, thoughtful responses, and not over-sharing to strangers. P22 put it as, \textit{``less of the ability to yell at people online and not be thoughtful.''} Life and dignity of the human person pertains to fostering authentic dialogue, which is bolstered by virtuous conversation, as opposed to fighting. Option for the poor and vulnerable pertains to user well-being, which is supported by virtuous conversation when people are not abused by harmful speech. Virtuous conversation was mentioned for the following patterns: Chats Over Feeds (N=9), Friends Over Followers (N=9), Moderated Mingling (N=13), Moderated Entry (N=10), No Lurking (N=2).

\textbf{Not exploitative} (\textit{life and dignity of the human person, option for the poor and vulnerable, care of God's creation}): Not exploitative means the application is not designed in a way that exploits the user for  profit, including keeping them online longer than necessary. In reference to the Notification Intentionality pattern, P10 said, \textit{``I think excessive notifications could start to manipulate our behavior...if we're always checking our notifications and not engaging in the world around us.''} Life and dignity of the human person and care of God's creation both relate to not requiring excess attention, and option for the poor and vulnerable relates to prioritizing user well-being over excess profits. Thus, these principles relate to \textit{not exploitative}. Not exploitative was mentioned for the following patterns: Chats Over Feeds (N=8), Friends Over Followers (N=1), Moderated Mingling (N=3), Clear Algorithmic Comprehension (N=5), No Lurking (N=3), and Notification Intentionality (N=21).

\textbf{Control} (\textit{subsidiarity}): This pertains to users having greater control over the content they see and the way online communities are governed. P15 said applications \textit{``should enable you to do what you want to do with [them].''} The principle of subsidiarity pertains to control at the lowest level and moderation, which relates to this participant-identified value of control. Control was mentioned for the following patterns: Chats Over Feeds (N=1), Moderated Mingling (N=5), Clear Algorithmic Comprehension (N=19), and  Notification Intentionality (N=5).

\textbf{Online Safety and Privacy} (\textit{life and dignity of the human person, option for the poor and vulnerable}): This value relates to users having greater privacy which prevents harassment, allowing for conversation in a more trusted environment, and keeping personal information secure. As an example, P20 noted that the Friends Over Followers pattern  \textit{``would minimize the dangers of social media...if you're mostly communicating with your friends, then the bullying is minimized, the creeps are minimized, all of that is minimized.''} Life and dignity of the human person calls for private spaces for greater moral exploration, and option for the poor and vulnerable calls for greater user well-being, both of which are embodied by online safety and privacy. Online safety and privacy was mentioned for the following patterns: Chats Over Feeds (N=2), Friends Over Followers (N=6), Clear Algorithmic Comprehension (N=2), Moderated Entry (N=4), and No Lurking (N=6).

\textbf{Transparency} (\textit{option for the poor and vulnerable}): Our participants want to understand why they are seeing the content they are seeing online. One reason is that it helps explain uncomfortably accurate recommendations: \textit{``I think having transparency is important so that way we can all have peace of mind about who is getting this information or why they're recommending things that are weirdly relevant to us''} (P6). Additionally, transparency can help users be more critical of their beliefs: \textit{``If you show people [why they're seeing certain content], then they think, do I agree with this philosophy?''} (P19). Because option for the poor and vulnerable called for transparent technologies, it pertains most to this value. Transparency was mentioned for the following patterns: Clear Algorithmic Comprehension (N=16), Moderated Entry (N=1), and No Lurking (N=1).

\textbf{Autonomy} (\textit{life and dignity of the human person; call to family, community, and participation; option for the poor and vulnerable}): This value pertains to users feeling they can participate in conversations freely and actively: \textit{``you're actually actively participating and talking with other people''} (P5). In this way it promotes the call to family, community, and participation which relates to forging stronger connections between people when users can participate actively in conversations. Participation in conversations also fosters the authentic dialogue called for in life and dignity of the human person. It also embodies option for the poor and vulnerable, which relates to user well-being, because people feel free to share their opinions without being harassed: \textit{``I know what topics are interesting to this group, and I know that I'm not going to be judged''} (P12). Autonomy was mentioned for the following patterns: Chats Over Feeds (N=3), Friends Over Followers (N=3), Moderated Mingling (N=1), Moderated Entry (N=2), and No Lurking (N=9).

We gave the design patterns embodiment scores for each principle of Catholic Social Teaching by adding together the number of times each of the values embodying that particular principle was mentioned for that pattern. For example, Moderated Entry had two participants mention autonomy and eleven participants mention personal connection, for a total \textit{call to family, community, and participation} embodiment score of thirteen. These embodiment scores will be used in Section~\ref{sec:overall} along with the violation scores calculated in Section~\ref{sec:violation} to compute a net embodiment score for each pattern. 

\subsubsection{Mapping the Participant-Identified Concerns to Violations of Catholic Social Teaching}\label{sec:violation}

\begin{figure*}
    \centering
    \includegraphics[width=\linewidth]{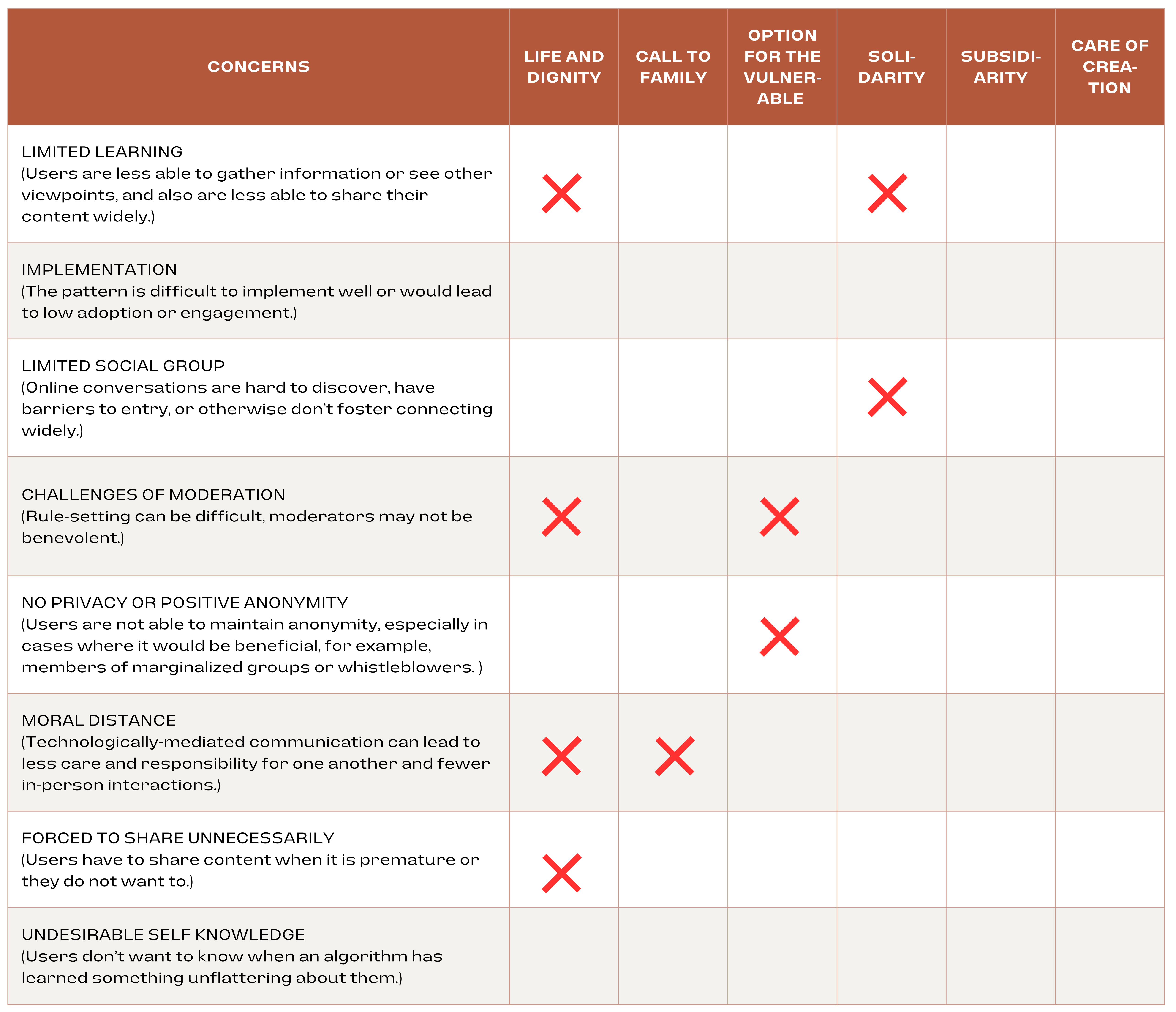}
    \caption{Concerns that our interview participants had with our design patterns, and which principles of Catholic Social Teaching they violate. Subsidiarity and care of God's creation are not violated by any of the participant-identified concerns. Life and dignity of the human person has the most violations.}
    \label{fig:downside_themes_to_cst}
        \Description{Comparison chart comparing the participant-identified concerns our design patterns embody with the principles of Catholic Social Teaching used to inspire the patterns. For each concern, the themes of Catholic Social Teaching it violates are identified with an X. The concerns listed down the left side are as follows. Limited learning (Users are less able to gather information or see other viewpoints, and also are less able to share their content widely.) Life and dignity and solidarity are X-ed. Implementation (The pattern is difficult to implement well or would lead to low adoption or engagement.) No Catholic Social Teaching principles are X-ed. Limited social group (Online conversations are hard to discover, have barriers to entry, or otherwise don’t foster connecting widely.) Solidarity is X-ed. Challenges of moderation (Rule-setting can be difficult, moderators may not be benevolent.) Life and dignity and option for the vulnerable are X-ed. No privacy or positive anonymity (Users are not able to maintain anonymity, especially in cases where it would be beneficial, for example, members of marginalized groups or whistleblowers.) Option for the vulnerable is X-ed. Moral distance (Technologically-mediated communication can lead to less care and responsibility for one another and less in-person interactions.) Life and dignity and call to family are X-ed. Forced to share unnecessarily (Users have to share content when it is premature or they do not want to.) Life and dignity is X-ed. Undesirable self knowledge (Users don’t want to know when an algorithm has learned something unflattering about them.) No Catholic Social Teaching principles are X-ed.}
\end{figure*}

We performed the same process for the violations of Catholic Social Teaching. We first considered the eight concerns highlighted by our participants and which principles of Catholic Social Teaching they violate, if any. This mapping can be seen in Figure~\ref{fig:downside_themes_to_cst}.\\

\textbf{Limited learning} (\textit{life and dignity of the human person, solidarity}): This concern embodies users being unable to gather information or see other viewpoints, and also being unable to share their content widely. P7 said, \textit{``you're not exposed to as much of what's out there, which can broaden your opinion''}, and P19 noted that, \textit{``it's harder to see public figures or people who are inspiring.''} Solidarity relates to engaging with global issues, and life and dignity of the human person relates to engaging in dialogue. Both of these are hindered when access to information is limited. Limited learning was mentioned for the following patterns: Chats Over Feeds (N=15), Friends Over Followers (N=14), Moderated Mingling (N=11), Clear Algorithmic Comprehension (N=11), Moderated Entry (N=3), No Lurking (N=19), and Notification Intentionality (N=6).

\textbf{Implementation:} This embodies the pattern being difficult to implement well, or concerns that the pattern would lead to low adoption, engagement, or revenue. For example, for Clear Algorithmic Comprehension, participants noted, \textit{``the idea is not bad, but I'm more concerned about the execution''} (P4), and \textit{``I don't have anything negative to say, other than it might impact revenue''} (P12). This concern does not violate any themes of CST because it is more about the implementation of the platform rather its usage. Implementation was mentioned for the following patterns: Chats Over Feeds (N=5), Friends Over Followers (N=2), Moderated Mingling (N=1), Clear Algorithmic Comprehension (N=7), Moderated Entry (N=5), No Lurking (N=3), and Notification Intentionality (N=14).

\textbf{Limited social group} (\textit{solidarity}): This embodies concerns that our patterns create online spaces that are hard to discover, join, or otherwise don’t enable connecting widely. P1 said, \textit{``Part of the point of social media is to connect with people you don't know.''} Solidarity pertains to connecting with the global human family, which is not possible when social groups are limited. Limited social group was mentioned for the following patterns: Chats Over Feeds (N=9), Friends Over Followers (N=10), Moderated Mingling (N=2), Moderated Entry (N=13), and No Lurking (N=1).

\textbf{Challenges of moderation} (\textit{life and dignity of the human person, option for the poor and vulnerable}): This concern encompasses difficulties of rule-setting in moderation and concerns that moderators may not be benevolent. P4 raised the concern, \textit{``what if people are trying to contribute to the conversation, but the community guidelines decide that's off topic or that's hate speech?''} P15 additionally raised a concern about \textit{``abuses of power''} from moderators. 
While these concerns are often mitigated by good moderators, the larger the group the more inevitable these issues become~\cite{gillespie2018custodians}. Life and dignity of the human person pertains to the ability to have authentic dialogue, and option for the poor and vulnerable pertains to user well-being, both of which could be violated if moderators are not benevolent or are not able to enforce good rules (either because of their own incompetence or because the group exists at too large a scale). Challenges of moderation was mentioned for the following patterns: Moderated Mingling (N=14), and Moderated Entry (N=4).

\textbf{No privacy or positive anonymity} (\textit{option for the poor and vulnerable}): This concern encompasses users not being able to maintain anonymity, especially in cases where it would be beneficial, for example, members of marginalized groups or whistleblowers. In response to Moderated Entry, P1 raised the concern, \textit{``for groups that are discriminated against, having their identity verified is potentially dangerous for them.''} Option for the poor and vulnerable encompasses prioritizing well-being and considering the needs of marginalized groups, both of which are violated with this concern. No privacy or positive anonymity was mentioned for: Moderated Entry (N=12), No Lurking (N=1).

\textbf{Moral distance} (\textit{life and dignity of the human person; call to family, commmunity, and participation}:) This concern encompasses the fear that technologically-mediated communication leads to less care and responsibility for one another (P4 said about Chats Over Feeds, \textit{``I can get away with saying things if I don't have to see their face or any impact''}) and fewer in-person interactions (P18 worried Friends Over Followers would create \textit{``an illusion of closeness''} and P20 worried Friends Over Followers \textit{``might encourage more online conversations instead of in-person''}). Life and dignity of the human person relates to fostering authentic dialogue, which is bolstered when people bring their whole selves to the conversation~\cite{conwill2024virtue}. The call to family, community, and participation relates to fostering stronger interpersonal connections. Thus, moral distance relates to violations of these principles. Moral distance was mentioned for the following patterns: Chats Over Feeds (N=6), Friends Over Followers (N=5), and Moderated Mingling (N=2).

\textbf{Forced to share unnecessarily} (\textit{life and dignity of the human person}): This concern encompasses users having to share content when it is premature or they do not want to. P1 expressed, \textit{``if my only way to read different points of view is to force myself into a conversation or argument that I am not educated enough to be a part of, that's just going to cause more toxicity.''} Life and dignity of the human person pertains to authentic dialogue, and this concern means dialogue is worse when people are forced to participate when they may not be ready to. Forced to share unnecessarily was mentioned only for No Lurking (N=4).

\textbf{Undesirable self-knowledge} This concern encompasses users not wanting to know when an algorithm has learned \textit{``something unflattering about [them]''} (P15). This concern does not correspond to violations of any principles of Catholic Social Teaching. Undesirable self-knowledge was only mentioned for Clear Algorithmic Comprehension (N=2).\\

\begin{figure*}
    \centering
    \includegraphics[scale=0.13]{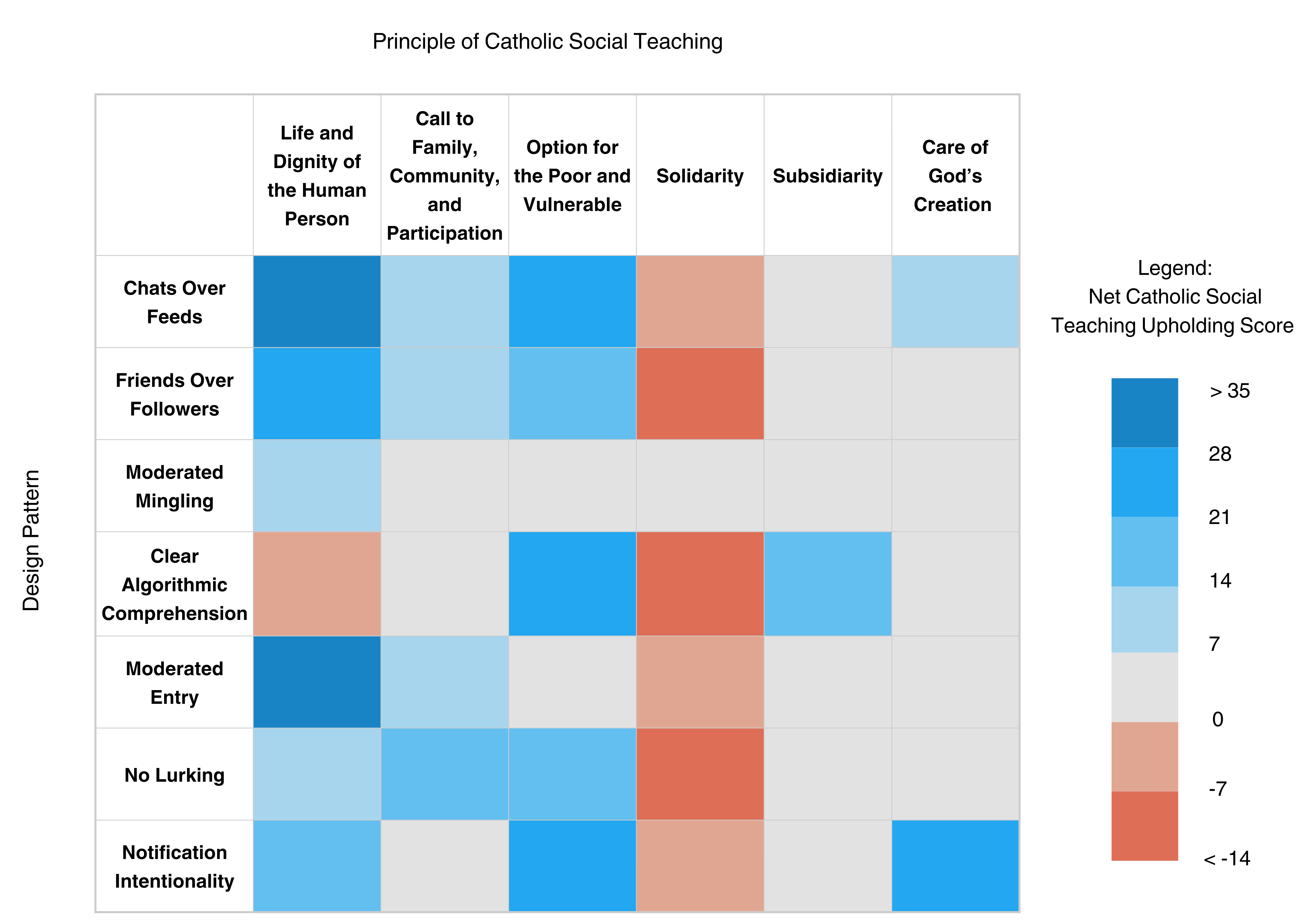}
    \caption{Heat map describing which principles of Catholic Social Teaching are embodied or violated by our design patterns. Most principles of Catholic Social Teaching are embodied by our patterns, although our patterns struggle to embody solidarity.}
    \Description{A heat map is shown with the design patterns on  the y axis, the Catholic Social Teaching principles on the x axis, and each square at the intersection of the design pattern and the Catholic Social Teaching principle is colored red, tan, or blue based on how well the design pattern embodies that principle. The legend shows that squares with an embodiment score below -7 are colored dark red, between -7 and 0 is light red, 0 and 7 is tan, 7 and 14 is light blue, 14 and 21 is a slightly darker shade of blue, 21 and 28 is blue, and 28 and above is dark blue. Chats Over Feeds has blue for all the principles except life and dignity of the human person is light red. The Call to Family, Community, and Participation has a mix of light blue and tan, option for the poor and vulnerable has mostly blue or light blue with tan on moderated entry and moderated mingling, solidarity has mostly light red or red with tan on moderated entry, subsidiarity has mostly tan except for light blue on clear algorithmic comprehension, and care of God's creation has mostly than with light blue on chats over feeds and blue on notification intentionality.}
    \label{fig:heatmap}
\end{figure*}

We gave the design patterns violation scores for each principle of Catholic Social Teaching by adding together the number of times each of the concerns violating that particular principle was mentioned for that pattern. For example, Chats Over Feeds had six participants mention increased moral distance and 15 participants mention limited learning, for a total \textit{life and dignity} violation score of 21. Because implementation concerns and undesirable self-knowledge did not correspond to violations of the principles of Catholic Social Teaching, we did not consider these concerns in the scores. The violation scores will be used in Section~\ref{sec:overall} along with the embodiment scores calculated in Section~\ref{sec:embodiment} to compute a net embodiment score for each pattern. 

\subsubsection{Overall, do the patterns embody or violate Catholic Social Teaching?}\label{sec:overall}
For each pattern and each principle of Catholic Social Teaching, we subtracted the violation score from the embodiment score to get the net score of how well that pattern embodies that Catholic Social Teaching principle. We then used these net scores to create a heat map representing how well each design pattern embodies each principle of Catholic Social Teaching. The heatmap can be seen in Figure \ref{fig:heatmap}.

We see from the heat map that in all of the design patterns, most of the principles of Catholic Social Teaching are embodied to some degree. \textit{Solidarity} is the main principle of Catholic Social Teaching that is not embodied in the patterns. \textit{Life and dignity of the human person} is not embodied in Clear Algorithmic Comprehension; this is because the ability to customize recommended content caused concern in our interview participants that users would not be exposed to a wide variety of ideas, which could hinder dialogue.

\textbf{Chats Over Feeds} was intended to foster \textit{life and dignity of the human person}, \textit{option for the poor and vulnerable}, and \textit{Care of God's Creation}. The heatmap indicates that these virtues, plus the \textit{call to family, community, and participation}, were embodied.

\textbf{Friends Over Followers} was intended to embody \textit{life and dignity of the human person} and \textit{subsidiarity}. The heat map indicates that the \textit{call to family, community, and participation} and the \textit{option for the poor and vulnerable} were also embodied. \textit{Subsidiarity} was not well-embodied, although it was not violated.

\textbf{Moderated Mingling} was intended to embody \textit{life and dignity of the human person}, \textit{option for the poor and vulnerable}, \textit{solidarity}, and \textit{subsidiarity}. The heat map indicates that while the other principles were not necessarily violated, \textit{life and dignity of the human person} was embodied the most.

\textbf{Clear Algorithmic Comprehension} was intended to embody \textit{option for the poor and vulnerable} and \textit{subsidiarity}. The heat map indicates it embodied these two the most compared to other principles.

\textbf{Moderated Entry} was intended to embody \textit{life and dignity of the human person} and \textit{subsidiarity}. It embodied \textit{life and dignity of the human person} well, and did not violate \textit{subsidiarity}, but embodied the \textit{call to family, community, and participation} more.

\textbf{No Lurking} was intended to embody the \textit{call to family, community, and participation}. The heat map indicates it additionally embodied \textit{life and dignity of the human person} and \textit{option for the poor and vulnerable}.

\begin{figure*}
    \centering
    \includegraphics[scale=0.22]{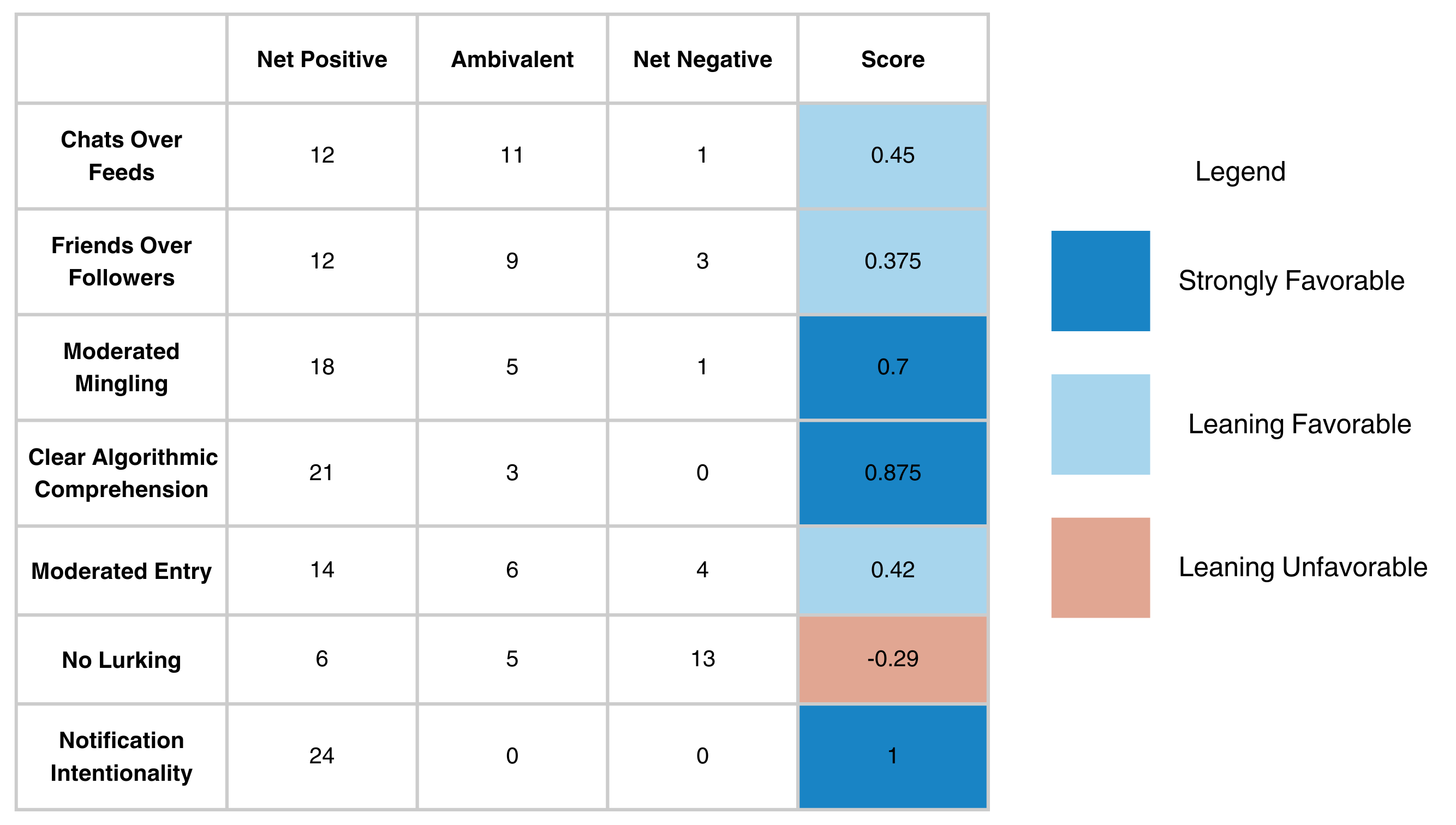}
    \caption{Total numbers of participants who described each pattern as a net positive, net negative, or ambivalent, and the score describing if the pattern was overall seen as strongly favorable, leaning favorable, leaning unfavorable, or strongly unfavorable by our participants. Only No Lurking was unfavorable with our participants, all other patterns were leaning favorable or strongly favorable.}
    \Description{A table with each of the design patterns and the number of net positive, ambivalent, and net negative responses, and their favorability scores. The legend indicates favorability scores in dark blue are strongly favorable, scores in light blue are leaning favorable, and scores in light red are leaning unfavorable. Chats Over Feeds had 12 net positive, 11 ambivalent, 1 net negative, and a favorability score of 0.45, colored in light blue for leaning favorable. Friends Over Followers had 12 net positive, 9 ambivalent, and 3 net negative for a favorability score of 0.337, colored in light blue for leaning favorable. Moderated Mingling had 18 net positive, 5 ambivalent, and 1 net negative for a favorability score of 0.7, colored in dark blue for strongly favorable. Clear algorithmic comprehension had 21 net positive, 3 ambivalent, and 0 net negative for a favorability score of 0.875, colored in dark blue. Moderated entry ahd 14 net positive, 6 ambivalent, and 4 net negative for a favorability score of 0.42, colored in light blue for leaning favorable. No lurking had a net positive of 6, ambivalent of 5, and net negative of 13, for a favorability score of -0.29, colored in light red. Notification intentionality and 24 net positive, 0 ambivalent, 0 net negative, for a favorability score of 1, colored in dark blue.}
    \label{fig:netposneg}
\end{figure*}

\textbf{Notification Intentionality} was intended to embody \textit{life and dignity of the human person}, \textit{option for the poor and vulnerable}, and \textit{care of God's creation}. The heat map indicates it embodied these three more than the other principles.

Thus, for \textit{\textbf{RQ2:} Does the proposed process work as intended, identifying design patterns that indeed embody the desired virtues?} we conclude that it does for the most part. Each pattern embodied most of the principles it was intended to embody, although in many cases extra principles were embodied, and some of the intended principles were not as well-embodied as expected. The \textit{call to family, community, and participation} and \textit{option for the poor and vulnerable} were often embodied when not intended, and \textit{subsidiarity} was often not as well-embodied as expected. As a whole, the patterns embody our chosen principles of Catholic Social Teaching well, and they especially embodied \textit{life and dignity of the human person}; \textit{option for the poor and vulnerable}; and the \textit{call to family, community, and participation}. However, they did not embody \textit{solidarity} well.

\subsection{Do people like the patterns? Would they use a social media designed with these patterns?}
We determined the answer to \textbf{RQ3} by considering how many participants rated each pattern as ``net positive,'' ``net negative,'' or ``net neutral.'' In the interviews ``net neutral'' was primarily interpreted as ``ambivalent'' by the participants; in other words, the pattern had both strong values and strong concerns such that the participant could not choose between positive and negative. Because of this, we will refer to ``net neutral'' responses as ``ambivalent'' in our results.

For each design pattern, we calculated a score to determine how favorable the pattern was across participant responses. The score was calculated by subtracting the number of net negative responses from the number of net positive responses and dividing that by the number of participants. Each pattern received a score from $-1$ to $1$, where $-1$ indicated unanimous disfavor, and $1$ indicated unanimous favor. Scores for each pattern can be seen in Figure \ref{fig:netposneg}.

\textbf{No Lurking} was the only pattern that leaned unfavorable, with limited learning (N=19) as the most commonly brought up concern. Many participants highlighted the benefits of lurking on the Internet to be able to gain information or learn about community norms before participating.

\textbf{Notification Intentionality} was unanimously favorable, and not exploitative (N=21) was the most commonly mentioned value: many participants noted that excess notifications are often intended to steal the user's attention and get them back on the app.

\textbf{Moderated Mingling} and \textbf{Clear Algorithmic Comprehension} were also strongly favorable. The most commonly mentioned values for Moderated Mingling were meaningful conversation (N=17) and virtuous conversation (N=13): our interview participants liked that moderation can help keep conversation respectful and on topic. The most commonly mentioned values for Clear Algorithmic Comprehension were control (N=19) and transparency (N=16).

\textbf{Chats Over Feeds}, \textbf{Friends Over Followers}, and \textbf{Moderated Entry} were leaning favorable but had some ambivalent and net negative responses. For both Chats Over Feeds and Friends Over Followers, the top values were meaningful conversation (N=17, N=10) and personal connection (N=14, N=14), and the top concerns were limited learning (N=15, N=14) and limited social group (N=9, N=10). Our participants appreciated the intentional conversation and connection that both patterns foster, but would not want this design pattern to dominate the social media landscape at the expense of the ability to connect with a wider group of people and learn about a wider variety of topics. The top values mentioned for Moderated Entry were meaningful conversation (N=12), personal connection (N=11), and virtuous conversation (N=10), and its most commonly mentioned concerns were limited social group (N=13) and no privacy or positive anonymity (N=12). Our participants valued that identity verification for group membership could lead to better connections and conversations, but also worried about exclusivity, undesirable breaches of privacy in the identity verification, and the inability to stay anonymous in situations where it could be beneficial.

Because all but one of the patterns were strongly favorable or leaning favorable, and no patterns were strongly unfavorable, we conclude that for \textit{\textbf{RQ3:} Are the design patterns identified by this process seen as desirable by technology users?} the answer is generally, yes.

\subsection{Do tech professionals think these patterns have a place in industry?}
Eight of the twenty-four participants whose experience related to product design/product management/human factors engineering, mobile development, or front-end engineering were asked additional questions about if our proposed catalog of design patterns should or would be adopted in industry. (Because our proposed design patterns were identified from popular applications, they are in a sense already adopted in industry. However, we often found them alongside less virtuous designs in the wild. Evaluating our proposed design patterns as a catalog emphasizes that we want to know if these patterns would be adopted in industry when divorced from the less virtuous designs found in the same application.) Every participant expressed that overall the patterns would have a positive effect on the human person if adopted, but five of the eight participants expressed doubts with the patterns being adopted in big tech because of how many of them resist extractive money-making designs. P19 said:

\textit{``For a lot of companies, making money is the number one priority. So they can look at these and and be like, `Oh that looks nice. But our Q2 goals tell us that we need to step it up on the user engagement,' so that goes.''}

Similarly, P24 expressed that tech companies would make use of these patterns only if they were in a financially secure enough place to do so. P20 noted that while big tech companies may not use the patterns as is because they wouldn't make enough money, they may incorporate elements of the patterns that allow them to orient their apps slightly more towards human flourishing without sacrificing profit too much.

On a more hopeful note, two participants expressed that while big tech may not adopt these patterns, smaller socially-conscious companies or nonprofits would value them. In fact, P13, who was a product designer for a Catholic company building a chat platform, noted that our design patterns resonated with the company's design priorities:

\textit{``Every single one was relevant. I could think of examples of decisions we're trying to make actively that relate to each of these. I think they fit in really closely [to our goals].''}

Three participants noted that each of our proposed patterns has trade-offs and are more or less appropriate for particular technologies depending on their goals or context. These participants believe our proposed design patterns could have a place in industry as long as their pros and cons are clearly articulated, as well as which types of technologies they are more appropriate for. P8 said, \textit{``I think that I would consider using them if I know right away what the pros and cons are.''} This should be mitigated by the \textit{consequences} dimension of the design patterns, which we did not show our interview participants to not bias their responses about the positive and negative impacts of each pattern.

Two of the eight participants who were asked about our proposed patterns' place in industry were atheist/agnostic. These two participants were additionally asked about how the Catholic Social Teaching principles inspiring the patterns resonated with their personal value systems. Both participants noted that the principles of Catholic Social Teaching resonated with their personal values despite not being Christian. One participant noted that from their knowledge of Catholicism, some of the specific details of how a Catholic may live out the Catholic Social Teaching principles may be different from how they would; for example, for a Catholic, the call to family, community, and participation may have a stronger emphasis on the nuclear family, whereas for this participant, they may put less of an emphasis on blood or legal relation. However, the over-arching ideas still resonated.

Thus, for \textit{\textbf{RQ4:} Would the catalog of design patterns created by this process be adopted in technology development?} we conclude that while big tech companies may not adopt these patterns, socially-conscious companies or individual programmers may adopt these patterns, regardless of their religious background.

\section{Discussion}
We discuss the design implications of our findings, adoption of our design patterns and the nature of the online spaces they would create, iterating on the design patterns for further value discovery, limitations, and opportunities for future work.

\subsection{Design Implications: The Importance of Subsidiarity}
Across all patterns, the top three most mentioned values were meaningful conversation, personal connection, and virtuous conversation. The fourth most mentioned value was not exploitative, and not exploitative was the most mentioned value for the two design patterns to receive the highest favorability scores from our participants (Clear Algorithmic Comprehension and Notification Intentionality). Taken together, this means that our participants value social media designs that promote authentic conversation and connection, and are not exploitative. In designing such technologies, we recommend designing for subsidiarity. Technology that upholds subsidiarity will foster either small-group interactions or moderated large-group interactions, and will give the user more control and customization over their experience. Small-group interactions or moderated large-group interactions will in turn promote more respectful and authentic conversations. They may also limit exploitative mechanisms from tech companies: exploitative mechanisms often rely on there being excessive amounts of content and advertising for the user to endlessly consume, which is limited in these designs. Allowing users to customize their experience more, including greater control over recommended content, the number of notifications one receives, and moderation guidelines, can also resist exploitative mechanisms. Although we are not the first to propose subsidiarity as a value that could promote better technologies~\cite{hasinoff2022scalability}, we believe subsidiarity should be a greater part of the conversation on how to build positive technology.

Four of the seven patterns we cataloged had subsidiarity as one of the underpinning principles of Catholic Social Teaching. These were Friends over Followers, Moderated Mingling, Clear Algorithmic Comprehension, and Moderated Entry. Interestingly, only Clear Algorithmic Comprehension embodied subsidiarity in the interview evaluations. In the case of our patterns, subsidiarity was a primary motivating factor in the design but wasn't fostered as much in the use of the technology, according to our participant-identified values. This indicates that when designing for virtue, there may be times where a particular virtue is designed for, but the resulting design gives rise to different virtues in its use. This is consistent with the idea in virtue ethics that the virtues are connected: possessing one virtue necessarily gives rise to other virtues~\cite{Budziszewski_2017}.

\subsection{Adoption of the Patterns}
Although our interview participants rated a majority of the patterns as having a net positive effect on the social media landscape, the participants who had industry experience in product, user experience, or front-end engineering roles expressed doubt that ``big tech'' companies would adopt these patterns due to the way they overwhelmingly push back against extractive designs.

We do not expect big tech companies to adopt our virtue ethics-based design process: that may be an impossible goal given their economic incentives. In fact, from the Industrial Revolution to the modern day, Catholic Social Teaching has continually criticized unchecked capitalism for its violation of human dignity. Catholic Social Teaching presents a human-centered economic vision that is antithetical to the hyper-competitive and efficiency-driven values of Silicon Valley today. Designing to align with the principles of Catholic Social Teaching would certainly clash with the goals of big tech. Instead, we propose a more radical vision in which individuals frustrated with the current state of the Internet move away from large tech conglomerates towards more community-based and open-source technology developed using our Virtue-Guided Technology Design process. The community-based and open-source vision of the Internet is consistent with the pre-Internet Bulletin Board Systems of the 1980s which media theorist Kevin Driscoll puts forth as an inspiration for a better social media landscape today~\cite{driscoll_modem_2022}.

Although these patterns were identified to be in accord with Catholic virtues, we believe they would be adopted by non-Catholics as well. The atheist and agnostic study participants expressed that the principles of Catholic Social Teaching underpinning these design patterns resonated with their personal value systems, which indicates that Catholic Social Teaching in particular could be easily brought into multicultural settings. This is fitting given that Catholic Social Teaching is not meant just for Catholics, but is intended to bring all people concerned with social justice into conversation. The design patterns proposed in this paper embody the principles of Catholic Social Teaching at a general enough level such that they resonated cross-culturally amongst our study participants, but tensions could arise if future patterns were created in a more specifically Catholic sense. To use the example provided by one of our participants, if a design pattern that upholds the \textit{call to family, community, and participation} over-emphasizes connecting with one's nuclear family, this pattern may not resonate with communities that focus less on the importance of blood or legal relation. When employing Virtue-Guided Technology Design, we encourage the consideration of the wider resonance of the chosen traditions' virtues across other value systems.

\subsection{The Vision Put Forth By These Patterns}
The vision put forth by our seven design patterns is a vision of intentional conversations, either in unmoderated small groups or in moderated large-group chat rooms. While participants generally liked this vision, as exhibited by many ``net positive'' responses to the patterns, they did not want this type of interaction to be the only type available. This is seen in the high number of ambivalent responses, especially for the Chats Over Feeds and Friends Over Followers patterns, and also through the concerns expressed. The concern mentioned the most across all patterns was limited learning, which encompasses not being able to see other viewpoints or gather information. The limited learning concern violated the principle of solidarity, which we found was the primary principle of Catholic Social Teaching that our patterns violated. Thus we see a dichotomy emerge between fostering intentional conversation and gathering information or seeing new ideas. How do we reconcile the fact that smaller groups and more guardrails can help to foster more intentional conversations, but these very same guardrails can curtail the spread of ideas? The Moderated Mingling pattern was intended to foster this large group sharing of ideas. However, it seems that participants did not find this to be a fully satisfying way of sharing ideas widely and they expressed doubts that moderation would foster a truly free sharing of ideas.

Because whether or not our participants liked the patterns was often context-dependent, the patterns would be improved with a ``context'' dimension that clearly explains for what types of technologies the pattern should be used, or to include this in the ``consequences'' dimension. Additionally, technologies should be intentionally designed for single specific contexts. If different designs are needed to do personal connection well versus doing idea-discovery well, it may be beneficial to split out these designs into different applications rather having platforms like Instagram that both foster connections with real-life friends and promote discovering new ideas through influencer content. This vision is in line with the principle of \textit{subsidiarity}, and is supported by previous research in HCI that shows that social media designs focusing on a single purpose help promote more authentic sharing online~\cite{kim2024sharing}. 

\subsection{Iterating on the Design Patterns for Further Value Discovery}
An accepted critique of Value Sensitive Design is Le Dantec's proposal to evolve it away from value prescription and towards value discovery~\cite{le2009values}. Our proposed Virtue-Guided Technology Design process appears to violate this critique because of the prescription of virtues embedded in the design patterns. However, this apparent violation of Le Dantec's proposal can be reconciled with another element of the virtue ethics tradition, as well as the existing culture surrounding design pattern documentation.

Specifically, the virtue ethics tradition holds that virtue is not learned and perfected in a single moment, but rather that someone striving for virtue must always iterate and continue to grow. We propose something similar with the design patterns we have put forth in this paper: we must continue to look for and document virtuous design patterns, and we can critique and discuss the ones we currently have. This is also consistent with the vision for software design patterns articulated in \textit{Design Patterns: Elements of Reusable Object-Oriented Software.} In the preface, the authors say they do not consider the collection of design patterns in the book to be complete and static: they welcome criticisms and additions~\cite{gamma1995pattern}.

This is especially important given that the translation of virtues to technical features is a subjective process. In this work, the authors relied on our own previous experience and engagement with Catholic Social Teaching when identifying its virtues in designs. The presence of bias from the researchers in their interpretation of the virtues does not render the work invalid: we have to start somewhere. Rather, this is an invitation for greater conversation surrounding the translation of these virtues into technology designs. This is especially important given the cultural diversity within Catholicism itself. Although the Catholic Church holds certain doctrines to be true, the way these doctrines are interpreted and lived out may vary across cultures. For example, Catholicism in the United States may look different from Latin American Catholicism, which may look different from African Catholicism. The design patterns put forth in this paper are merely a starting point based on our interpretations. In addition to our hope that researchers will use our Virtue-Guided Technology Design process with other faith traditions, we also hope that other researchers from around the world will use this process to further engage with Catholic Social Teaching, helping to draw out more of its richness because of their unique experiences and perspectives.

\subsection{Limitations and Opportunities for Future Research}
One limitation of our study was that the patterns were primarily evaluated by technology professionals with a Catholic or other Christian background. The Christian participants gave a wide variety of feedback and none of the atheist/agnostic and Muslim participants gave outlying responses compared to the rest of the group, suggesting that the feedback we received from Christians was not favorably biased. Nevertheless, we recommend future work with a more diverse set of participants. Additionally, all of our interview participants resided in the United States. It would be beneficial to repeat this process with interview participants from other countries, especially non-Western countries, to see if the patterns are still favorable.

The design patterns we documented for social media were a proof of concept for a virtue-based design process. Because they were inspired by Catholic Social Teaching, they reflect one particular set of virtues in their design. Our participants generally liked the designs but wanted there to be additional options, especially given that our designs prioritized intentional conversation at the expense of information-gathering. This leaves room for future work to study what connecting widely and information-gathering would look like when done virtuously, especially minimizing exploitative mechanisms and impersonal or disrespectful conversations. One way to foster creativity in this process is to repeat it with virtues from other traditions. Are there insights we can gain from considering Confucian, Buddhist, Hindu, or Islamic virtues that are lost in the western perspective that could help us to reconcile the dichotomy between intentional conversation and exposure to new ideas? Considering other virtue traditions would also address the call for more ethical pluralism in HCI~\cite{ahmed2022situating, rifat2023many}.

Additionally, contrasting the Catholic Social Teaching-abiding design patterns with design patterns that uphold other faith-based or ethical traditions could have strengthened the study. Most of our study participants approved of the Catholic Social Teaching-abiding patterns. But, our argument would have been strengthened by seeing design patterns upholding other faith traditions accepted by our participants and design patterns with no ethical backing rejected. We recommend future work exploring this further.

As discussed in the previous section, our bias with our own interpretation of the principles of Catholic Social Teaching could be viewed as a limitation. We recommended that researchers and designers with different backgrounds and perspectives continue to iterate on the design patterns we put forth. In addition to this, we recommend future work that investigates how designers perceive and define values, especially in light of their familiarity with particular value systems.

Finally, one limitation of our approach is that other aspects of the technology development process, including design processes and company organization, are left out when focusing only on design patterns. We did not consider the Catholic Social Teaching principles \textit{dignity of work and rights of workers} and \textit{rights and responsibilities} in our approach because they pertain more to how the company is run than to the design of the technology itself. Additionally, the \textit{option for the poor and vulnerable} included ``prioritize the needs of marginalized groups in design'' which could entail incorporating the community-based design practices advocated by the \textit{Design Justice}~\cite{Costanza-Chock_2020} movement into the design process. We recommend future work considering how virtue can be incorporated into other aspects of how companies are run.

\section{Conclusion}
We proposed an approach for cataloging design patterns inspired by virtue ethics. When applying this approach to identify design patterns for social media grounded by Catholic Social Teaching, we found that the virtues inspiring the patterns were for the most part embodied. This supports the validity of our proposed Virtue-Guided Technology Design process. We also found that our participants liked the design patterns, regardless of religious adherence. This indicates that the virtues of Catholic Social Teaching resonate across worldviews, and bodes well for virtues from other traditions resonating widely. We propose that repeating this process with virtues from other traditions could help to generate additional virtuous design patterns for social media, and could possibly help to creatively address the dichotomy between fostering intentional conversation and the ability to gather information and see new ideas that emerged from our patterns. We propose a radical new vision of the Internet where socially conscious individuals resist against the exploitation of big tech through community-based and open-source technologies built with virtue-oriented design patterns.

%%
%% The acknowledgments section is defined using the "acks" environment
%% (and NOT an unnumbered section). This ensures the proper
%% identification of the section in the article metadata, and the
%% consistent spelling of the heading.
\begin{acks}
This work was supported by the Responsible Computing Challenge, a partnership of Omidyar Network, Mozilla, Schmidt Futures, Craig Newmark Philanthropies and Mellon Foundation. Any opinions, findings, conclusions, or recommendations expressed in this material are those of the authors and do not necessarily reflect the views of our sponsors.
\end{acks}

%%
%% The next two lines define the bibliography style to be used, and
%% the bibliography file.
\bibliographystyle{ACM-Reference-Format}
\bibliography{sample-base}

%%
%% If your work has an appendix, this is the place to put it.
\appendix
\section{Appendix}
Illustrations of the technical inquiry process for the other six design patterns whose illustrations were not included in the main body of this paper are found on the following pages.

\begin{figure*}
    \centering
    \includegraphics[width=0.8\linewidth]{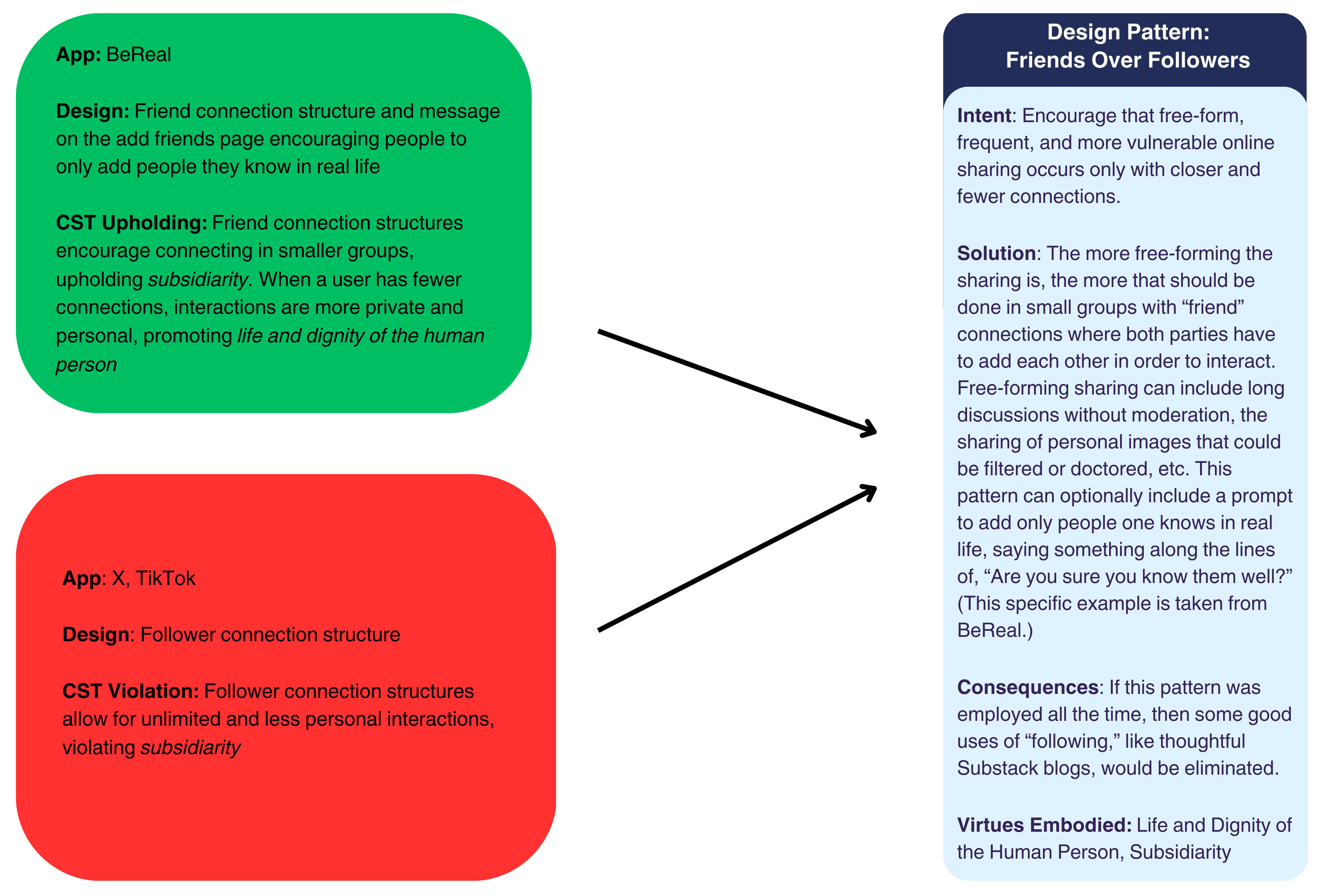}
    \caption{Technical Inquiry process for the Friends Over Followers pattern.}
    \Description{On the top there is a green bubble. It includes the following text: App: BeReal. Design: Friend connection structure and message on the add friends page encouraging people to only add people they know in real life. CST Upholding: Friend connection structures encourage connecting in smaller groups, upholding subsidiarity. When a user has fewer connections, interactions are more private and personal, promoting life and dignity of the human person.On the bottom there is a red bubble. It includes the following text: App: X, TikTok. Design: Follower connection structure. CST Violation: Follower connection structures allow for unlimited and less personal interactions, violating subsidiarity The two bubbles have arrows to a blue box that says: Design Pattern: Friends Over Followers. Intent: Encourage that free-form, frequent, and more vulnerable online sharing occur only with closer and fewer connections. Solution: The more free-forming the sharing is, the more that should be done in small groups with “friend” connections where both parties have to add each other in order to interact. Free-forming sharing can include long discussions without moderation, the sharing of personal images that could be filtered or doctored, etc. This pattern can optionally include a prompt to add only people one knows in real life, saying something along the lines of, “Are you sure you know them well?” (This specific example is taken from BeReal.) Consequences: If this pattern was employed all the time, then some good uses of “following,” like thoughtful Substack blogs, would be eliminated. Virtues Embodied: Life and Dignity of the Human Person, Subsidiarity}
\end{figure*}

\begin{figure*}
    \centering
    \includegraphics[width=\linewidth]{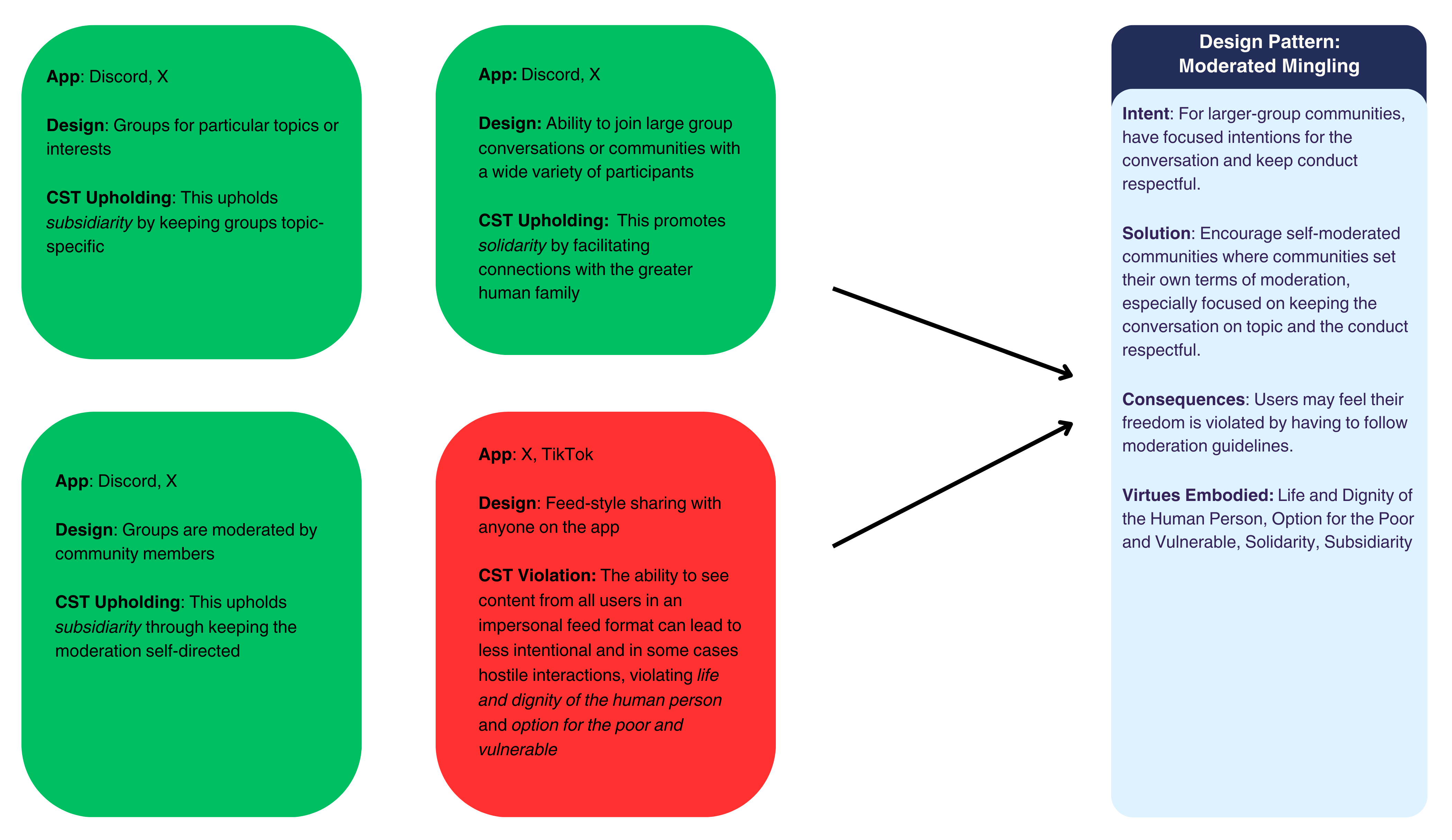}
    \caption{Technical Inquiry process for the Moderated Mingling pattern.}
    \Description{On the top there are two green bubbles and on the bottom there is one green bubble and one red bubble. The top left green bubble has the following text: App: Discord, X. Design: Groups for particular topics or interests. CST Upholding: This upholds subsidiarity by keeping groups topic-specific. The top right bubble has the following text: App: Discord, X. Design: Ability to join large group conversations or communities with a wide variety of participants. CST Upholding:  This promotes solidarity by facilitating connections with the greater human family. The bottom left green bubble has the following text: App: Discord, X. Design: Groups are moderated by community members. CST Upholding: This upholds subsidiarity through keeping the moderation self-directed. The bottom right red bubble has the following text: App: X, TikTok. Design: Feed-style sharing with anyone on the app. CST Violation: The ability to see content from all users in an impersonal feed format can lead to less intentional and in some cases hostile interactions, violating life and dignity of the human person and option for the poor and vulnerable. The bubbles have arrows to a blue box that says: Design Pattern: Moderated Mingling. Intent: For larger-group communities, have focused intentions for the conversation and keep conduct respectful. Solution: Encourage self-moderated communities where communities set their own terms of moderation, especially focused on keeping the conversation on topic and the conduct respectful. Consequences: Users may feel their freedom is violated by having to follow moderation guidelines. Virtues Embodied: Life and Dignity of the Human Person, Option for the Poor and Vulnerable, Solidarity, Subsidiarity.
}
\end{figure*}

\begin{figure*}
    \centering
    \includegraphics[width=0.7\linewidth]{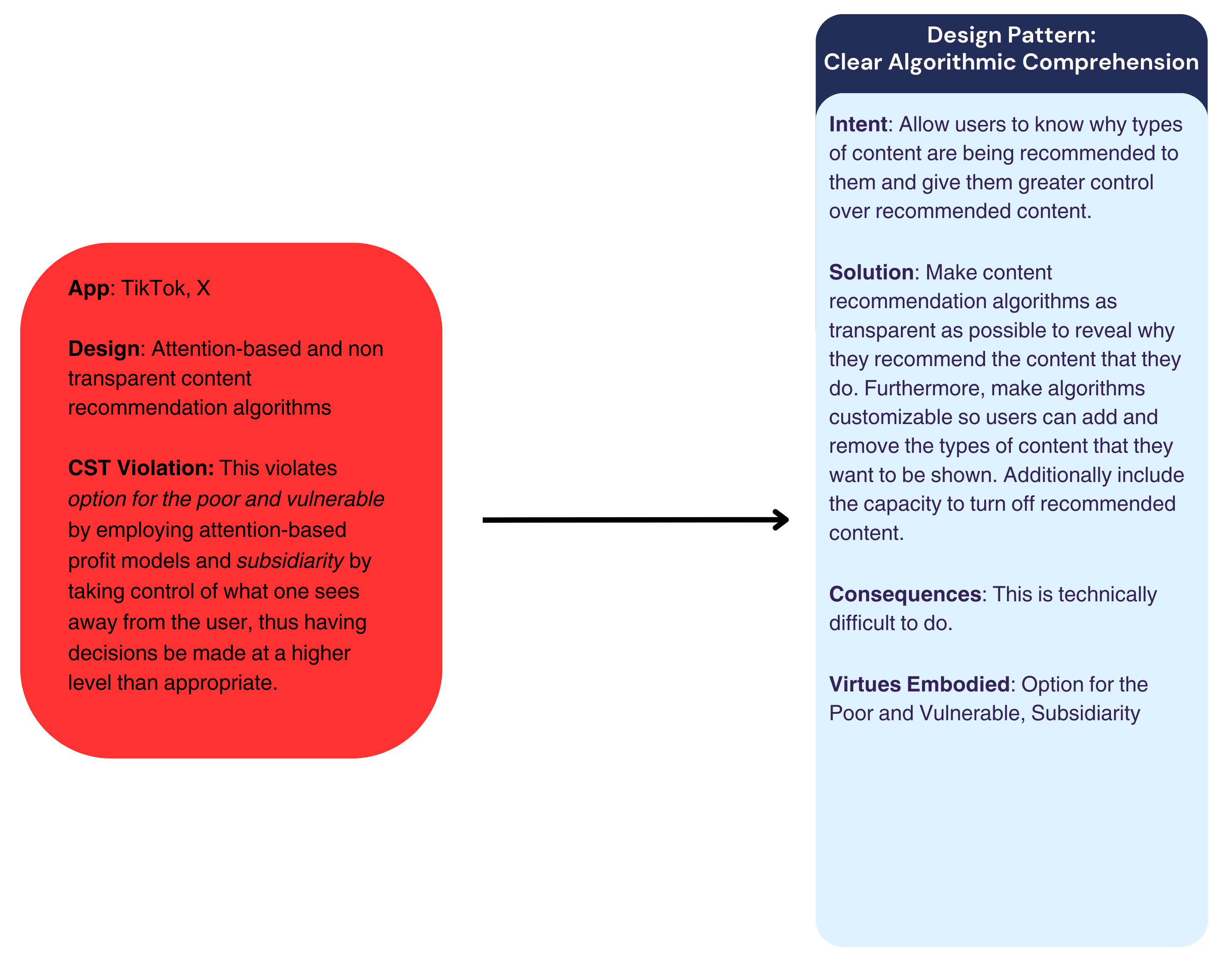}
    \caption{Technical Inquiry process for the Clear Algorithmic Comprehension pattern.}
    \Description{There is a red bubble with the following text: App: TikTok, X. Design: Attention-based and non transparent content recommendation algorithms. CST Violation: This violates option for the poor and vulnerable by employing attention-based profit models and subsidiarity by taking control of what one sees away from the user, thus having decisions be made at a higher level than appropriate. There is an arrow from the red bubble to a blue box with the following text: Design Pattern: Clear Algorithmic Comprehension. Intent: Allow users to know why types of content are being recommended to them and give them greater control over recommended content. Solution: Make content recommendation algorithms as transparent as possible to reveal why they recommend the content that they do. Furthermore, make algorithms customizable so users can add and remove the types of content that they want to be shown. Additionally include the capacity to turn off recommended content. Consequences: This is technically difficult to do. Virtues Embodied: Option for the Poor and Vulnerable, Subsidiarity.
}
\end{figure*}

\begin{figure*}
    \centering
    \includegraphics[width=\linewidth]{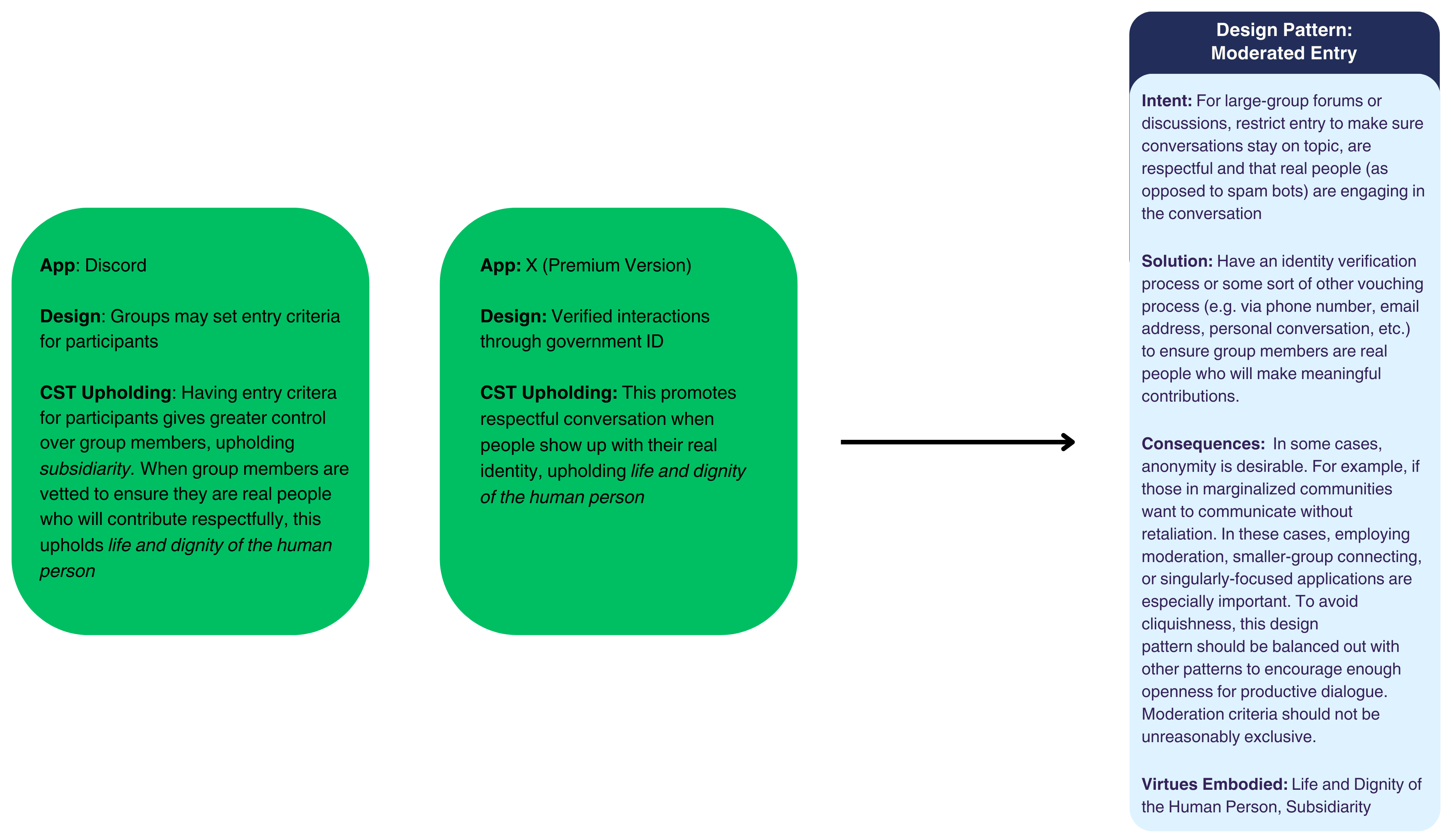}
    \caption{Technical Inquiry process for the Moderated Entry pattern.}
    \Description{There are two green bubbles. The green bubble on the left has the following text: App: Discord. Design: Groups may set entry criteria for participants. CST Upholding: Having entry critera for participants gives greater control over group members, upholding subsidiarity. When group members are vetted to ensure they are real people who will contribute respectfully, this upholds life and dignity of the human person. The green bubble on the right has the following text: App: X (Premium Version). Design: Verified interactions through government ID. CST Upholding: This promotes respectful conversation when people show up with their real identity, upholding life and dignity of the human person. There is an arrow pointing from the green bubbles to a blue box with the following text: Design Pattern: Moderated Entry. Intent: For large-group forums or discussions, restrict entry to make sure conversations stay on topic, are respectful and that real people (as opposed to spam bots) are engaging in the conversation. Solution: Have an identity verification process or some sort of other vouching process (e.g. via phone number, email address, personal conversation, etc.) to ensure group members are real people who will make meaningful contributions. Consequences:  In some cases, anonymity is desirable. For example, if those in marginalized communities want to communicate without retaliation. In these cases, employing moderation, smaller-group connecting, or singularly-focused applications are especially important. To avoid cliquishness, this design pattern should be balanced out with other patterns to encourage enough openness for productive dialogue. Moderation criteria should not be unreasonably exclusive. Virtues Embodied: Life and Dignity of the Human Person, Subsidiarity
 }
\end{figure*}

\begin{figure*}
    \centering
    \includegraphics[width=0.7\linewidth]{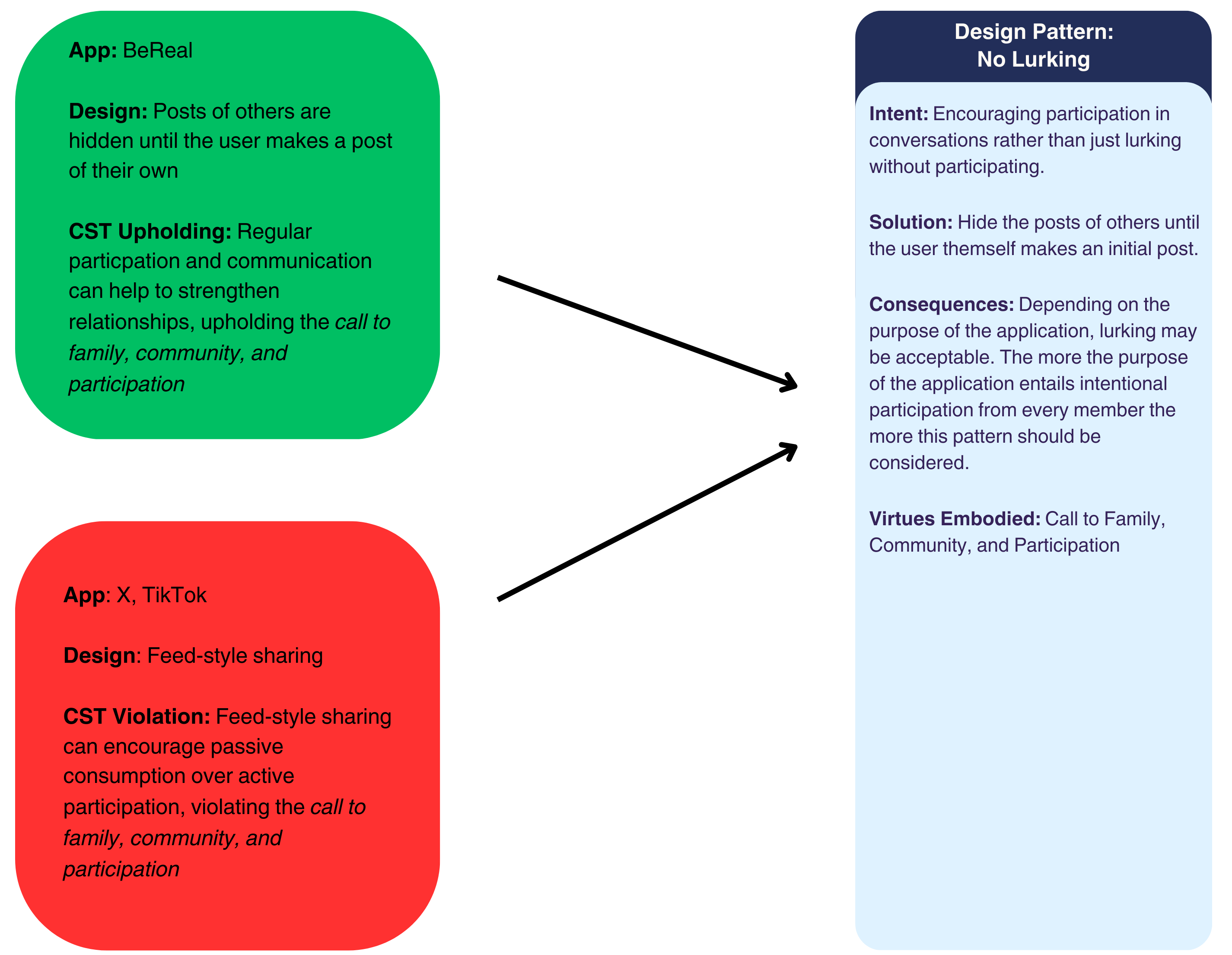}
    \caption{Technical Inquiry process for the No Lurking pattern.}
    \Description{There is a green bubble on top of a red bubble. The green bubble has the following text: App: BeReal. Design: Posts of others are hidden until the user makes a post of their own. CST Upholding: Regular particpation and communication can help to strengthen relationships, upholding the call to family, community, and participation. The red bubble has the following text: App: X, TikTok. Design: Feed-style sharing. CST Violation: Feed-style sharing can encourage passive consumption over active participation, violating the call to family, community, and participation. There are arrows coming from the bubbles to a blue box with the following text: Design Pattern: No Lurking. Intent: Encouraging participation in conversations rather than just lurking without participating. Solution: Hide the posts of others until the user themselves makes an initial post. Consequences: Depending on the purpose of the application, lurking may be acceptable. The more the purpose of the application entails intentional participation from every member the more this pattern should be considered. Virtues Embodied: Call to Family, Community, and Participation}
\end{figure*}

\begin{figure*}
    \centering
    \includegraphics[width=0.7\linewidth]{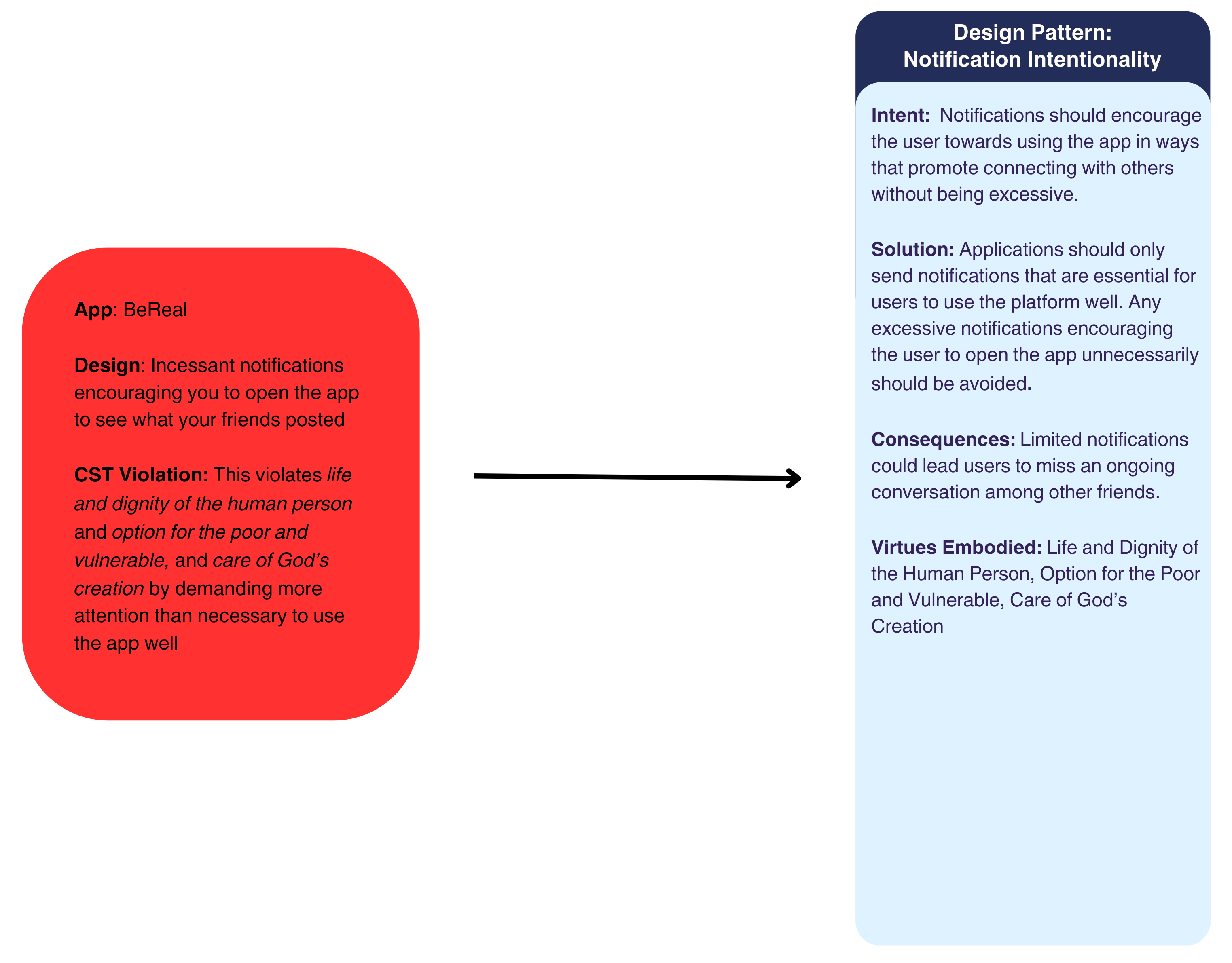}
    \caption{Technical Inquiry process for the Notification Intentionality pattern.}
    \Description{There is a red bubble with the following text: App: BeReal. Design: Incessant notifications encouraging you to open the app to see what your friends posted. CST Violation: This violates life and dignity of the human person and option for the poor and vulnerable, and care of God’s creation by demanding more attention than necessary to use the app well. There is an arrow pointing from the red bubble to a blue box with the following text: Design Pattern: Notification Intentionality. Intent:  Notifications should encourage the user towards using the app in ways that promote connecting with others without being excessive. Solution: Applications should only send notifications that are essential for users to use the platform well. Any excessive notifications encouraging the user to open the app unnecessarily should be avoided. Consequences: Limited notifications could lead users to miss an ongoing conversation among other friends. Virtues Embodied: Life and Dignity of the Human Person, Option for the Poor and Vulnerable, Care of God’s Creation}
\end{figure*}

\end{document}

%% file: participant_table.tex
\begin{table*}[htb]
\caption{\textbf{Participant Demographics}}
\label{tab:participant_demographics}
\begin{tabular}{|clllll|}
\hline
\textbf{ID\#} & \textbf{Gender} & \textbf{Race/Ethnicity} & \textbf{Religion} & \textbf{Tech Work Experience}           & \textbf{Domain}   \\ \hline
P1            & Man             & White                   & Catholic          & AI research, Software engineering       & Academia/Industry \\ \hline
P2            & Man             & White                   & Catholic          & AI research/development                 & Industry          \\ \hline
P3            & Woman           & White                   & Catholic          & Software engineer                       & Industry          \\ \hline
P4            & Woman           & White                   & Catholic          & Software engineer                       & Industry          \\ \hline
P5            & Man             & White                   & Catholic          & AI research/development                 & Industry          \\ \hline
P6            & Woman           & White                   & Protestant        & Security research                       & Academia          \\ \hline
P7            & Woman           & White                   & Catholic          & Mobile developer                        & Industry          \\ \hline
P8            & Woman           & White                   & Catholic          & UX design, HCI research                 & Academia/Industry \\ \hline
P9            & Woman           & Middle Eastern          & Muslim            & AI research                             & Academia          \\ \hline
P10           & Man             & White                   & Catholic          & Math/Computer Science research          & Industry          \\ \hline
P11           & Man             & White                   & Catholic          & Software engineer                       & Industry          \\ \hline
P12           & Man             & White                   & Catholic          & AI research                             & Academia          \\ \hline
P13           & Woman           & White                   & Catholic          & Product designer                        & Industry          \\ \hline
P14           & Woman           & White                   & Protestant        & HCI research                            & Academia          \\ \hline
P15           & Man             & White                   & Catholic          & ML engineering, Product design          & Industry          \\ \hline
P16           & Man             & White                   & Catholic          & AI researcher                           & Academia/Industry \\ \hline
P17           & Man             & Asian                   & Catholic          & Front end engineer                      & Industry          \\ \hline
P18           & Man             & Black                   & Catholic          & Software engineer/Data scientist        & Industry          \\ \hline
P19           & Nonbinary       & Asian                   & Agnostic          & HCI research, Human factors engineer    & Academia/Industry \\ \hline
P20           & Woman           & Hispanic/Latino         & Catholic          & Mobile developer                        & Industry          \\ \hline
P21           & Man             & White                   & Catholic          & Software engineer                       & Industry          \\ \hline
P22           & Woman           & White                   & Catholic          & Software engineer                       & Industry          \\ \hline
P23           & Man             & Mixed Race              & Protestant        & Product management                      & Industry          \\ \hline
P24           & Man             & Asian                   & Atheist           & HCI researcher, AI research/development & Academia/Industry \\ \hline
\end{tabular}
\end{table*}